\documentclass[twocolumn]{aastex631}

\usepackage{amsmath}
\usepackage{graphicx}% Include figure files
\usepackage{dcolumn}% Align table columns on decimal point
\usepackage{bm}% bold math
\usepackage{xcolor}
\usepackage{hyperref}% add hypertext capabilities
\usepackage[normalem]{ulem}

\newcommand{\vx}{\mathbf x}
\newcommand{\vu}{\mathbf u}
\newcommand{\los}{\mathbf{\hat{n}}}

\newcommand{\vb}{\mathbf b}

\newcommand{\reff}{\rm ref}
\newcommand{\nuref}{\nu_{\reff}}
\newcommand{\relB}{B^{\rm rel}}
\newcommand{\indB}{\bar{B}}

\newcommand{\rev}[1]{{{#1}}}

\begin{document}

\title{Correlation-based Beam Calibration of 21cm Intensity Mapping}

\correspondingauthor{Xin Wang}
\email{wangxin35@mail.sysu.edu.cn}

\author{Jiacheng Ding}
\affiliation{School of Physics and Astronomy, Sun Yat-Sen University, 
\\ No.2 Daxue Road, Zhuhai 519082, China}
 
\author{Xin Wang}
\affiliation{School of Physics and Astronomy, Sun Yat-Sen University, 
\\ No.2 Daxue Road, Zhuhai 519082, China}
\affiliation{CSST Science Center for the Guangdong-Hong Kong-Macau Greater Bay Area, SYSU, China}

\author{Ue-Li Pen}
\affiliation{Institute of Astronomy and Astrophysics, Academia Sinica, Astronomy-Mathematics Building, 
\\ No.1, Sec.4, Roosevelt Road, Taipei 10617, Taiwan}
\affiliation{Canadian Institute for Theoretical Astrophysics, 60 St. George Street, Toronto, ON M5S 3H8, Canada}
\affiliation{Canadian Institute for Advanced Research, 180 Dundas St West, Toronto, ON M5G 1Z8, Canada}
\affiliation{Dunlap Institute for Astronomy and Astrophysics, University of Toronto, 
\\ 50 St George Street, Toronto, ON M5S 3H4, Canada}
\affiliation{Perimeter Institute of Theoretical Physics, 31 Caroline Street North, Waterloo, ON N2L 2Y5, Canada}

\author{Xiao-Dong Li}
\affiliation{School of Physics and Astronomy, Sun Yat-Sen University, 
\\ No.2 Daxue Road, Zhuhai 519082, China}
\affiliation{CSST Science Center for the Guangdong-Hong Kong-Macau Greater Bay Area, SYSU, China}

\begin{abstract}
Foreground removal presents a significant obstacle in both current and forthcoming intensity mapping surveys. While numerous techniques have been developed that show promise in simulated datasets, their efficacy often diminishes when applied to real-world data. A primary issue is the frequency-dependent variations in the instrumental response. In this paper, we propose a novel approach utilizing the internal cross-correlation among different frequencies to calibrate the beam's frequency fluctuations. Using a simulated dataset that incorporates frequency-dependent random fluctuations into the beam model, we illustrate that our method can achieve considerable improvements over traditional techniques. Our results represent a step forward in enhancing the precision and reliability of foreground removal in intensity mapping surveys. 
\end{abstract}


\section{\label{sec:intro} Introduction}
The large-scale structure of the Universe holds invaluable information, providing insights into various aspects such as galaxy evolution, fundamental physics including dark matter, dark energy, alternative theories of gravity, and the early Universe, etc. For over four decades, galaxy surveys have played an important role in enhancing our understanding of the Universe, including the 2dF Galaxy Redshift Survey\footnote{http://www.2dfgrs.net/} ~\cite{colless20012df}, the 6dF Galaxy Survey\footnote{http://www.6dfgs.net/}~\cite{jones20046df}, the WiggleZ Dark Energy Survey\footnote{https://wigglez.swin.edu.au/} ~\cite{drinkwater2010wigglez}, and SDSS-III BOSS\footnote{https://www.sdss3.org/}~\cite{anderson2014clustering}, DES\footnote{https://www.darkenergysurvey.org/}~\cite{dark2016dark}, DESI\footnote{https://www.desi.lbl.gov/}~\cite{collaboration2023early}, and LSST\footnote{https://www.lsst.org/}~\cite{ivezic2019lsst} in the future. In recent years, 21cm intensity mapping has emerged as a alternative technique to probe the large-scale structure \citep{scott199021}. This method, focusing on the 21cm line emission from neutral hydrogen (HI), differs from conventional galaxy surveys that observe individual galaxies. Instead, 21cm intensity mapping captures the collective emission from multiple galaxies, mapping the intensity fluctuation without resolving individual galaxies. This efficient approach enables us to rapidly map matter distribution over large volumes of Universe. Significant achievements have been made in this area over the past decade. These include the successful cross-correlation of 21cm maps with galaxy samples~\citep{masui2013measurement, wolz2022h, cunnington2023h} and the measurement of the auto-power spectrum using interferometer~\citep{abdurashidova2022first,paul2023first}. Consequently, numerous ongoing surveys and forthcoming next-generation radio telescopes (SKA\footnote{https://www.skao.int/en/}, HERA\footnote{https://reionization.org/}~\citep{deboer2017hydrogen}) are poised to significantly enhance the detection capabilities, and will eventually lead to cosmological constraints using this technique.

Despite these advancements, significant challenges remain, particularly with foreground contamination. Being several orders of magnitude brighter than the cosmological signal, these foregrounds substantially hinder progress in auto-correlation measurements, especially when compared to cross-correlation with galaxy surveys -- a milestone achieved over a decade ago. Recent breakthrough in detecting the auto-power spectrum using the MeerKAT interferometry mode and foreground avoidance techniques~\citep{paul2023first} marks significant progress. However, the long baselines of the MeerKAT telescope are confined to observing only smaller scales, which limits their ability in deriving meaningful cosmological constraints. 

Many different techniques have been devised to overcome this challenge, including methods like spectral fitting, Principal Component Analysis (PCA) \citep{masui2013measurement,switzer2013determination,switzer2015interpreting,bigot2015simulations}, Independent Component Analysis (ICA) \citep{chapman2012foreground,wolz2014effect}, and the Karhunen-Loève transform (KL) \citep{PhysRevD.105.083503} and other machine learning based methods  \citep{mangena2020constraining,makinen2021deep21,wadekar2021hinet,villanueva2021removing} etc. However, while many of these methods demonstrate reasonable effectiveness in extracting signals from idealized, simulated datasets, real-world complexities often hinder their practical application.

One complexity arises from the chromaticity of the instrument. In interferometric observations, the chromatic response causes a mixing of low-wavelength foreground modes with small-scale HI signal within a wedge-like region, further restricting the Fourier space where foreground avoidance techniques can be applied. While this effect is less a issue in telescopes operating in single-dish mode, chromaticity can still be a significant factor. This is largely due to the inherent imperfections in any instrument we construct, which introduce a multiplicative, frequency-dependent noise into the response beam. The convolution of this noisy beam with the observed sky can significantly degrade the effectiveness of most, if not all, existing foreground removal methods. This includes techniques ranging from blind PCA/ICA and parametric fitting to others that necessitate knowledge of the instrumental response, such as the Karhunen-Loève m-modes and various machine learning-based methods.

At the first glance, the most straightforward solution to mitigate beam variation appears to be simply improving the calibration or measurement accuracy of the beam. Indeed, this approach has been widely pursued, with significant efforts dedicated to improving beam understanding through methods such as drone measurement \citep{chang2015beam} and holographic measurements \citep{hunter2011holographic,iheanetu2019primary}. In this paper, however, we introduce a novel approach to address this issue. Our core concept is that it is the relative beam differences across frequencies that primarily influence the success of foreground removal. An unknown, frequency-independent deviation in the beam only introduces an additional window function to the detected signal, with minimal impact on the efficiency of the foreground removal process. Therefore, we propose a Correlation-Based Beam Calibration method (CBC), designed to internally calibrate the frequency variations among different frequency channels. 

The structure of this paper is as follows: Section II provides a concise overview of our method, outlining the fundamental concept. Section III delves into the detailed methodology. Section IV describes the process of generating the simulated data, followed by the presentation of our results in Section V. The paper concludes with a discussion in Section VI. In this paper, we assume a fiducial cosmology with parameters of ($\Omega_{\rm m}\!=\!0.3$, $\Omega_{\rm b}\!=\!0.049$, $\Omega_{\Lambda}\!=\!0.7$, $w\!=\!-1.0$,  $\sigma_{8}\!=\!0.8$, $h\!=\!0.67$, $n_{s}=0.96$)

\section{Instrumental Chromaticity and Foreground Removal}
\label{sec:method_review}
To demonstrate the fundamental principle of our method, let us begin by presuming that the observed data have undergone processing up to the map-making stage, yielding an intensity map. This map, represented as the temperature $T_{\nu}(\los)$, is a function of the line-of-sight (LoS) direction $\los$ and the frequency $\nu$, which is composed of foregrounds and cosmological signal
\begin{eqnarray}
T_{\nu}(\los)= T^{\rm HI}_{\nu}(\los)+ T^{\rm FG}_{\nu}(\los).
\end{eqnarray}

To proceed, we make the basic assumption that the observed sky map is a convolution of the true sky $T_{\nu}(\los)$ with a frequency-dependent power beam $B_{\nu}(\los)$
\begin{eqnarray}
\label{eqn:conv}
\widetilde{T}_{\nu}(\los) = \int d\los^{\prime} B_{\nu}(\los-\los^{\prime})
T_{\nu}(\los^{\prime}) + n_{\nu}(\los),
\end{eqnarray}
where $\widetilde{T}_{\nu}(\los)$ represents the observed sky temperature, with the tilde indicating the quantity as post-instrumentally `observed'. 

As will be seen, this assumption is essential here, as our methodology relies on filtering within the Fourier space. For many radio telescopes, this assumption holds reasonably true, although the observational strategy or subsequent data analysis procedures might introduce some position-dependent response functions. A case in point is the 19-beam L-band receivers of the Five-hundred-meter Aperture Spherical radio Telescope (FAST). In its drift scan mode \citep{zhang2019status,Hu2021MNRAS,li2023fast}, disparate sky regions may be observed by different subsets of these 19 beams. Consequently, without proper compensation during the map-making process, any disparities among the receivers might lead to a violation of our assumption.

\begin{figure}
\centering
\includegraphics[width=0.48\textwidth]{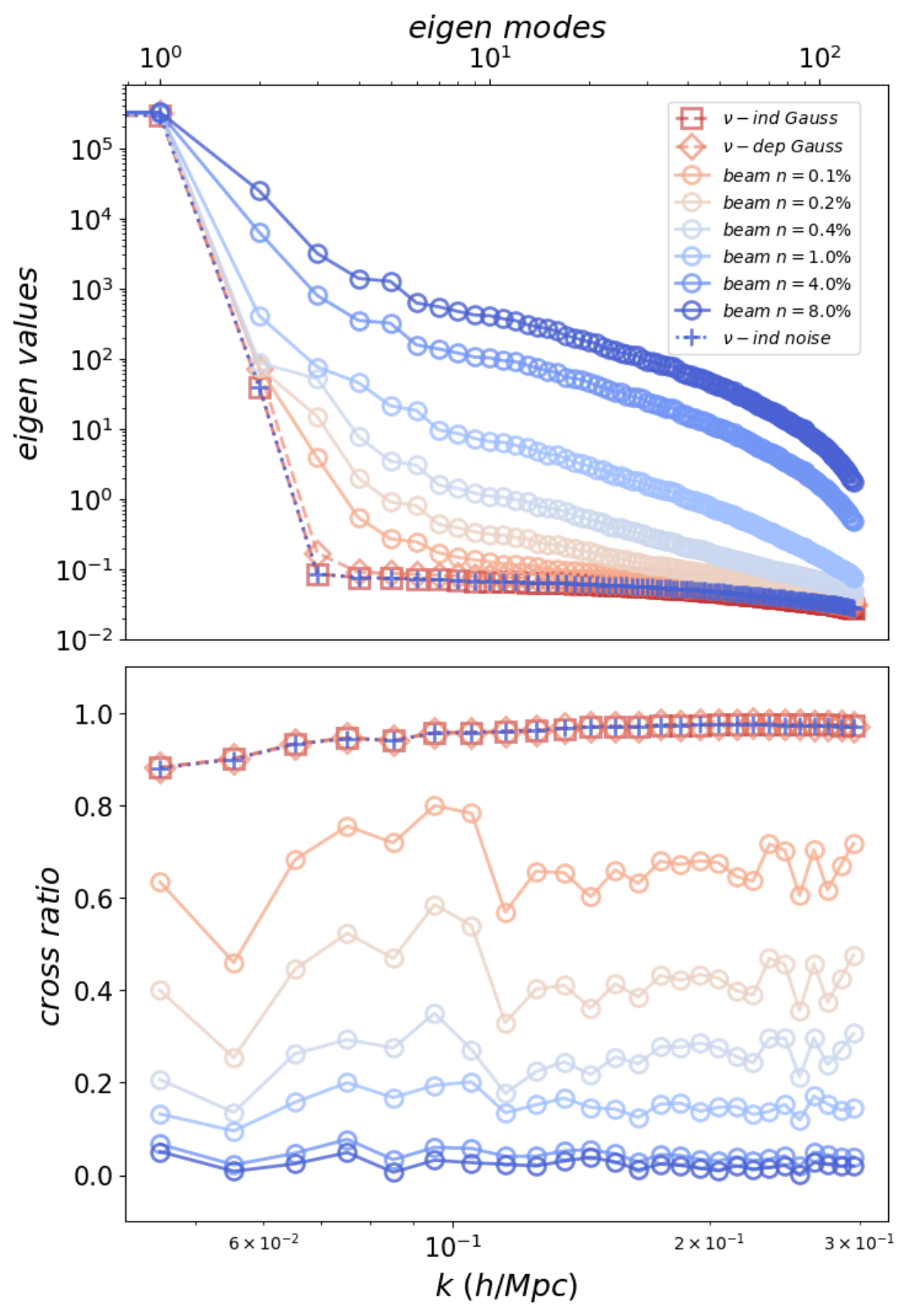}
\caption{\label{fig:intro_demo}
Demonstration of SVD foreground removal with frequency dependent beam noise. The {\it upper} panel displays the eigenvalues for various beam models. Here `$\nu$-ind Gauss' refers to the frequency-independent Gaussian beam, while `$\nu$-dep Gauss' represents the frequency-dependent Gaussian model. The notation `$\nu$-ind noise' is used for the noisy Gaussian beam without frequency variation, and finally the `beam $n=\varepsilon\%$' denotes the frequency-dependent random noise with noise amplitude of $\varepsilon\%$. As demonstrated, a noisy yet frequency-independent beam does not affect the eigenvalues. In contrast, the influence of a frequency-dependent noisy beam is heavily dependent on the amplitude of the noise. The {\it lower} panel illustrates the cross-correlation ratio with the HI signals after removing the first $3$ SVD modes. The color and line scheme here mirrors that of the upper panel. 
Clearly, the frequency-dependent beam noises introduce confusion and result in a larger foreground residual, leading to a substantially lower cross-correlation.  As shown, even with just a $0.1\%$ level of beam noise, there is a notable reduction in the cross-correlation ratio between the cleaned map and the true HI signal.}
\end{figure}

In the Fourier space, Equation (\ref{eqn:conv}) is expressed as 
\begin{eqnarray}
\label{eqn:Tobs}
\widetilde{T}_{\nu}(\vu) = B_{\nu}(\vu)\,T_{\nu}(\vu) + n_{\nu}(\vu),  
\end{eqnarray}
where $\vu = \nu \vb /c$ is the Fourier conjugate of sky coordinates $\los$. Here $c$ is the speed of light and $\vb$ is the spatial separation between elements of the instrument. In an ideal radio telescope observation, the beam function scales linearly with frequency. In Fourier space, this implies that the dependency on $\nu$ is expressed through the relation
\begin{eqnarray}
\label{eqn:uvscaling}
 \vu = \nu \vb /c.
\end{eqnarray}

In reality, no instrument is flawless, and these imperfections could significantly influence the performance of foreground removal techniques. Many existing methods for foreground subtraction rely on having at least some knowledge of the beam characteristics. An imperfect or noisy telescope beam can compromise the accuracy of these techniques. Here, the `imperfect beam' refers to any deviation from the ideal antenna response pattern, which can stem from a range of issues including manufacturing inaccuracies, mechanical structural deformations of the antenna, errors in feed positioning, and frequency-dependent responses of electronic components, etc. Environmental conditions such as temperature fluctuations and wind can induce additional deformations. While these factors indeed influence the beam, their temporal variability also challenges our assumption of a consistent convolution process. As such, they fall outside the scope of this paper.

To gain a clearer understanding of how frequency-dependent variations in the beam response can cause distortion and potential intermixing of foreground and cosmological signals, it is instructive to examine specific cleaning techniques. For our analysis, existing foreground removal methods can be divided into two categories. The first category includes methods that explicitly require knowledge of the beam function. This group encompasses forward-modeling methods \citep{Bernardi_2011MNRAS,Sullivan2012ApJ}, the m-mode formalism which relies on the input beam function to construct a signal-to-contaminant-ratio ordered Karhunen–Loève basis \citep{Shaw_2014ApJ}, and various semi-blind methods that assume prior knowledge of both the foreground and the instrument \citep{Zuo2023ApJ}. Additionally, this category includes machine learning-based approaches \citep{gillet2019deep,la2019machine,makinen2021deep21,ni2022eliminating, gheller2022convolutional,shi202321}. For these methods, a precise understanding of the instrumental response is essential. The second category includes methods that do not explicitly require an instrumental response. However, they can still be significantly affected by a noisy beam. For instance, some techniques that rely on the spectrum's smoothness, such as parametric and certain non-parametric fitting methods, are particularly vulnerable. A noisy beam disrupts the underlying assumption of smoothness, rendering these techniques less effective or inapplicable.

Another type of method within this category is the blind methods, notably PCA and ICA, as detailed in various studies \citep{masui2013measurement, alonso2015blind, bigot2015simulations, stone2002independent, alonso2015blind, tharwat2021independent}. These techniques are generally more adept at handling instrumental systematics than parametric fitting approaches and are widely employed in real data analysis, making them the primary focus of our paper. In these methods, the data cube $\mathbf{D} = \mathbf{F}+\mathbf{S}+\mathbf{N}$ is interpreted as a combination of the foregrounds $\mathbf{F}$, the 21 cm signal $\mathbf{S}$, and instrumental noise $\mathbf{N}$. Here, $\mathbf{D}$ is expressed as a two-dimensional matrix in the notation $\mathbb{R}^{n \times p}$, where $n$ is the number of frequency channels and $p$ the number of pixels. The goal is to derive a cleaned map $\mathbf{T}^{\rm cl}$ by excluding the foreground elements from the observed data. This is achieved as follows:
\begin{eqnarray}
\label{eqn:pcasub}
\mathbf{T}^{\rm cl} = \mathbf{D} - \left( \mathbf{U} \mathbf{I}_{\rm m} \mathbf{U}^{\rm T} \right) \mathbf{D},
\end{eqnarray}
where $\mathbf{U}$ represents the components matrix, which could be either the most dominant modes (PCA) or the most non-Gaussian modes (ICA). The first $\rm m$-components are selected with matrix $\mathbf{I}_{\rm m}$, signifying approximate foregrounds modeling as $\mathbf{F} \approx \left( \mathbf{U}\mathbf{I}_{\rm m}\mathbf{U}^{\rm T} \right) \mathbf{D}$. However, chromatic fluctuations in the instrument may modify the decomposed modes, leading to a confusion between the 21cm signal and beam fluctuations, thereby reducing the efficacy of these blind methods. To illustrate this, consider a simplified noisy beam model. In this model, three-dimensional random noise $\delta B_{\nu}(\vu)$ is added in the frequency-$\vu$ space to a Gaussian beam  $B^{G}_{\nu}(\vu)$:
\begin{eqnarray}
    B_{\nu}(\vu) &=& B^{G}_{\nu}(\vu) + \delta B_{\nu}(\vu) = B^{G}_{\nu}(\vu) (1+\epsilon_{\nu}(\vu))
\end{eqnarray}
where $\epsilon_{\nu}(\vu)$ represents a uniformly distributed random number. The specifics of this model are further elaborated in section \ref{subsec:simbeam}.

In Figure (\ref{fig:intro_demo}), we illustrate the impact of a noisy beam on PCA foreground removal. The upper panel illustrates the eigenvalues resulting from the SVD decomposition, where a rapid decrease in eigenvalues indicates that the spectral-smooth foreground can be effectively captured by the initial few modes. As shown, with a frequency-independent beam (marked by {\it red squares}), the first four eigen-modes sufficiently describe the foregrounds over six orders of magnitude. This sets the baseline result for the most ideal scenario. Similarly, a frequency rescaling of the Gaussian beam (see Eq. \ref{eqn:uvscaling}) does not significantly alter the results (shown as {\it red diamons}). More crucially, a frequency-independent noisy beam, where $\delta B_{\nu}(\vu)$ varies only in the $uv$ plane but not in the frequency direction (indicated by {\it blue crosses}), aligns closely with our ideal beam model. This observation is further supported in the lower panel, which presents the cross-correlation ratio between the true cosmological signal and the `cleaned' map after the subtraction of three SVD modes.

Conversely, with a increasing level of the frequency-dependent beam noise $\epsilon_{\nu}$, the eigenvalues descend more gradually, indicating a more complex data structure. As a result, more modes must be removed to reduce the map amplitude, leading to a more significant loss of the signal. At the same time, the cross-correlation ratio between the cleaned map (with three subtracted SVD modes) and the HI map starts to decline, a consequence of increased residual power. Notably, even a beam noise level of $0.1\%$ can substantially reduce the cross-correlation ratio, dropping it from near $1$ to about $0.6$. This highlights the substantial impact of beam noise on effective foreground removal.

\section{Correlation-based Beam Calibration}
\label{sec:method}
The above discussion clearly demonstrates that a frequency-independent fluctuations in the beam pattern have a minimal impact on foreground cleaning. This observation highlights the importance of understanding the relative beam variation across frequencies, rather than focusing solely on accurately measuring its absolute pattern. We would like to remind readers that this concept is not entirely new. The traditional SVD method (denoted as `TM') has previously sought to convolve the observed dataset with a common beam function, albeit under the assumption that the beam is well described by a Gaussian function \citep{masui2013measurement}
\begin{eqnarray}
\label{eqn:Gaussian_beam_rel-filter}
f^{\rm G}_{\nu} (\vu) &=& \; \exp\left [-\frac{(\gamma \sigma^2_{\rm ref} - \sigma^2_{\nu} ) u^2 }{2} \right]  \nonumber\\
&=& \frac{B^G_{\nuref}}{B^G_{\nu}} 
\exp\left[-\frac{(\gamma -1) \sigma_{\rm ref}^2 u^2}{2} \right ] .
\end{eqnarray}
Here $\sigma(\nu)$ represents the size of the Gaussian beam at frequency $\nu$, with $\sigma_{\rm ref}$ being the beam width at the reference frequency which is also the maximum value of $\sigma(\nu)$. The scaling factor, $\gamma$, is selected from a range of values, such as $1.2$ in MeerKAT study \citep{cunnington2023h} or $1.4$ in GBT measurement \citep{condon2016essential}, to further smooth out any noise arising from this process. It becomes evident that applying the filter $f^{G}$ to the observed map results in a map convolved with the same Gaussian beam at all frequencies
\begin{eqnarray}
{T}_{\nu} (\vu) &=& f^{G}_{\nu} (\vu ) \widetilde{T}_{\nu}(\vu) = f^{G}_{\nu}(\vu)\,B^{G}_{\nu}(\vu)\,T_{\nu}(\vu) \nonumber \\
&=& B^{G}_{\nuref} \exp\left[-\frac{(\gamma -1) \sigma_{\rm ref}^2 u^2}{2} \right ] T_{\nu}(\vu). 
\end{eqnarray}
Subsequently, in this section, we will expand on this concept and introduce our correlation-based approach, designed to effectively calibrate for the frequency-dependent variations in the beam pattern.

\subsection{Frequency-dependent Beam Fluctuation}
Generally, a realistic beam function $B_{\nu}(\vu)$ typically has a complex functional form, showing variations in both $\vu$ and $\nu$. As demonstrated in the previous section, the efficacy of foreground removal depends on the beam's relative response. This prompts the decomposition of the beam into two distinct components:
\begin{align}
B_{\nu}(\vu) = \indB(\vu) \relB_{\nu}(\vu),
\end{align}
where $\indB$ denotes the frequency-independent part, aligned with the beam at the reference frequency, i.e., $\indB = B_{\nuref}$. On the other hand, the frequency-dependent relative beam $\relB$ is defined as $\relB_{\nu} = B_{\nu}/\indB$. In the following, the primary objective of our method is to determine the relative beam $\relB_{\nu}$ with existing data.

To achieve this, we aim to develop a filtering function designed to uniformize the beam response across all frequencies
\begin{eqnarray}
\label{eqn:relfilt}
f_{\nuref\nu}(\vu) = \frac{B_{\nuref}(\vu)}{B_{\nu}(\vu)} = \frac{1}{\relB_{\nu}(\vu)} 
\end{eqnarray}
\rev{
Consequently, applying this filter $f_{\nuref\nu}(\vu)$ to the observed map $\widetilde{T}_{\nu}(\vu)$ 
\begin{eqnarray}
 f_{\nuref\nu}(\vu)\,\widetilde{T}_{\nu}(\vu)   % \\ \nonumber
 &=& f_{\nuref\nu}(\vu)\,B_{\nu}(\vu)\,T_{\nu}(\vu)  \\ \nonumber
 &=& \bar{B}(\vu)\,T_{\nu}(\vu). 
\end{eqnarray}
results in a map $\bar{B}(\vu)\,T_{\nu}(\vu)$ that is convolved with the same reference beam $B_{\nuref}$ at every frequencies.
}

\begin{figure}
\centering
\includegraphics[width=0.5\textwidth]{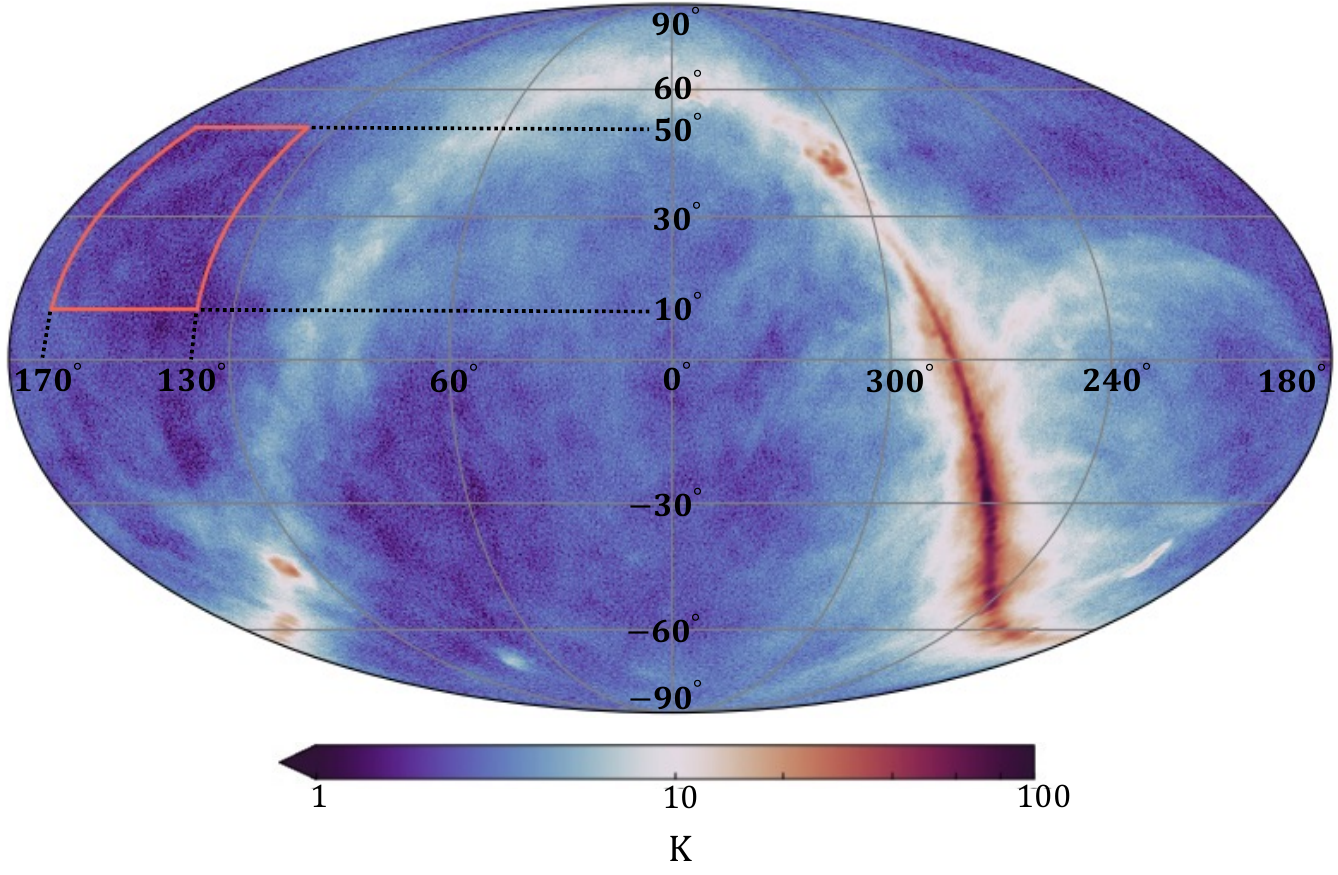}
\caption{ \label{fig:The_selected_sky-region}
The illustration depicts our sky model at a frequency of $800.2\,{\rm MHz}$. The highlighted region, which spans $130^{\circ} \le {\rm RA} \le 170^{\circ}$ and $10^{\circ} \le {\rm DEC} \le 50^{\circ}$ represents the area chosen for our simulated dataset.}
\end{figure}

\subsection{Correlation-based Beam Calibration}

\rev{Therefore, the core idea of our algorithm revolves around obtaining the relative beam $\relB_{\nu}$ through internal calibration. 
First, let us re-express the temperature map at a given frequency $\nu$ and $\vu$ cell as the product of the frequency-dependent relative beam $\relB_{\nu}(\vu)$ and the remaining component}
\begin{eqnarray}
\widetilde{T}_{\nu}(\vu) = \relB_{\nu}(\vu) \indB(\vu)T_{\nu}
=  \relB_{\nu}(\vu)\bar{T}_{\nu}(\vu) .
\end{eqnarray}
\rev{ Here we have ignored the noise term $n_{\nu}(\vu)$ and absorbed the frequency-independent beam in the temperature map and define }
\begin{eqnarray}
\label{eqn:Tbar}
\bar{T}_{\nu}=\indB(\vu) T_{\nu}. 
\end{eqnarray}
To proceed, we aim to construct a filter denoted as $F_{\nuref\nu}(\vu)$. Applying this filter to observed temperature  $\widetilde{T}_{\nu}(\vu)$ at each frequency channel effectively inverts the relative beam, yielding an estimator for $\bar{T}_{\nu}$
\begin{eqnarray}
\label{eqn:TWF_filter}
\bar{T}^{\rm WF}_{\nu}(\vu) = 
 F_{\nuref\nu}(\vu)\,\widetilde{T}_{\nu}(\vu). 
\end{eqnarray}
Consequently, our goal is to design the filter $F_{\nuref\nu}(\vu)$  so that it functions similarly to the relative filter $f_{\nuref\nu}(\vu)$ as described in Eq.~(\ref{eqn:relfilt}).
This could be achieved via the minimal variance estimator, derived by examining the variance 
\begin{eqnarray}
\label{eqn:r2u}
 r^2(\vu) &=& \left \langle\left(\bar{T}^{\rm WF}_{\nu}-\widetilde{T}_{\nuref} \right)
  \left (\bar{T}^{\rm WF}_{\nu}-\widetilde{T}_{\nuref}\right)\right \rangle (\vu) ~~~
\end{eqnarray}
where $\langle \cdots \rangle$ presents the averaging within specific grid cell in $\vu$. Minimizing $r^2 (\vu)$ leads to the well-known Wiener filter (WF) \footnote{hence this explains the meaning of the superscript ${\rm WF}$ in Eq. (\ref{eqn:TWF_filter}) }
\begin{eqnarray}
\label{eqn:wienerfilter}
F_{\nuref\nu}(\vu)
&=& \frac{\left\langle \widetilde{T}_{\nu}\,\widetilde{T}_{\nuref} \right\rangle}
         {\left\langle \widetilde{T}_{\nu}\,\widetilde{T}_{\nu}    \right\rangle} (\vu)
=   \frac{C^{\widetilde{T}}_{\nu \nuref}} {{C^{\widetilde{T}}_{\nu \nu}}} (\vu)
\end{eqnarray}
Here, we have introduced the correlation function $C^{\widetilde{T}}_{\nu_1 \nu_2}$, defined as
\begin{eqnarray}
C^{\widetilde{T}}_{\nu_1\nu_2} = \left\langle \widetilde{T}_{\nu_1}\,\widetilde{T}_{\nu_2} \right\rangle.
\end{eqnarray}
Substituting Eq.~(\ref{eqn:Tobs}) of the observed temperature map $\widetilde{T}$ and temporarily neglecting the noise matrix $N_{\nu_1\nu_2}=\langle n_{\nu_1}  n_{\nu_2}\rangle$, we obtain
\begin{eqnarray}
\label{eqn:Cobsexp}
C^{\widetilde{T}}_{\nu_1 \nu_2}(\vu) &=& \left\langle B_{\nu_{1}}\,T_{\nu_{1}}\,B_{\nu_{2}}\,T_{\nu_{2}}\right\rangle(\vu)     \nonumber \\
&=& \left\langle \relB_{\nu_1}\,\relB_{\nu_2}\, \bar{B}^{2}\,T_{\nu_{1}}\,T_{\nu_{2}} \right\rangle(\vu) .
\end{eqnarray}
It is apparent that
\begin{eqnarray}
\label{eqn:Ffilter}
F_{\nuref\nu}(\vu) &=& \frac{1}{\left\langle \relB_{\nu}(\vu) \right\rangle} \, F^{T}_{\nuref\nu}(\vu), 
\end{eqnarray}
where the additional intrinsic temperature factor $F^{T}_{\nuref\nu}(\vu)$ is defined as
\begin{eqnarray}
\label{eqn:F_T_filter}
F^{T}_{\nuref\nu}(\vu) 
&\equiv& \frac{ C^{T}_{\nu\nuref}(\vu) }{ C^{T}_{\nu\nu}(\vu)} .
\end{eqnarray}
\rev{It is apparent that if we ignore the frequency variation of the temperature map, causing the factor $F^{T}_{\nuref\nu}(\vu)$ to become unity, the constructed filter simply represents the average value of the relative beam across each given $\vu$ cell. Therefore, $F_{\nuref\nu}(\vu)$ is indeed the filter that meets our requirements. 
However, in reality, the image of the foreground varies with frequency, thus introducing an additional undetermined function on top of the relative beam. Various strategies can be employed to mitigate its influence, including using a foreground model as prior information. As we will discuss in detail in Section \ref{sec:results}, the impact of this term is not significant. }

\rev{Furthermore, observational data contains instrumental noise (Equation \ref{eqn:Tobs}), making the filter described in Equation (\ref{eqn:Ffilter}) usually sub-optimal for analyzing real data. 
Various methods can be adopted to effectively isolate the beam variations in such situation. We will explore these approaches in more detail and discuss their applications to real data analysis in a separate paper.
In this work, however, it is worth mentioning that a straightforward solution involves considering the full correlation matrix $C^{\widetilde{T}}_{\nu_i\nu_j}(\vu)$ across all frequencies 
\begin{eqnarray}
C^{\widetilde{T}}_{\nu_i\nu_j}(\vu) 
&=& \left\langle \widetilde{T}_{\nu_i}(\vu)\widetilde{T}_{\nu_j}(\vu) \right\rangle.
\end{eqnarray}
Here the observed temperature map $\widetilde{T}_{\nu_i}(\vu) $ is a combination of sky $\relB_{\nu}(\vu)\bar{T}_{\nu}(\vu)$ and noise $n_{\nu}(\vu)$. Assuming that the amplitude of the sky is much larger than instrumental noise, we can then perform the SVD of matrix $\bm{C}^{\widetilde{T}}(\vu)$
\begin{eqnarray}
\bm{C}^{\widetilde{T}}(\vu) = \bm{U} \bm{\Sigma} \bm{V}^{\dagger},
\end{eqnarray}
where the diagonal matrix ${\bm \Sigma} = {\rm diag}(\lambda_i)$ store all eigen-values in decending order, and $\bm{U}$ (and $\bm{V}$) lists the corresponding eigen-vectors. 
As long as the sky, primarily comprised of the foreground, dominates the matrix $C^{\widetilde{T}}_{\nu_i\nu_j}(\vu)$, one could approximate the observed temperature with the value of the first eigen-mode.
Therefore, in a similar approach to constructing the filter in Eq.~(\ref{eqn:Ffilter}), we can define the filter using only the first eigenvector $U^{(1)}_{\nu}(\vu)$, which can be expressed as:
\begin{eqnarray}
\label{eqn:Filt_obs}
F_{\nuref\nu}(\vu)&=&\frac{U_{\nuref}^{(1)} (\vu) }{U^{(1)}_{\nu} (\vu)}.
\end{eqnarray} 
The only underlying assumption here is the dominance of the sky temperature relative to the instrumental noise, which is achievable with sufficient integration time.
Consequently, Eq. (\ref{eqn:Filt_obs}), representing the ratio between the first eigen-modes at two different frequencies, characterizes the desired filter with reduced noise influence. 
Therefore, this approach has the advantage of requiring minimal information about the noise covariance matrix. 
}

\section{Simulation Data}
\label{sec:simdata}
In this section, we will provide a detailed overview of the technical specifics of our simulated data set, covering both the sky and beam models.

\begin{figure}
    \centering
    \includegraphics[width=0.5\textwidth]{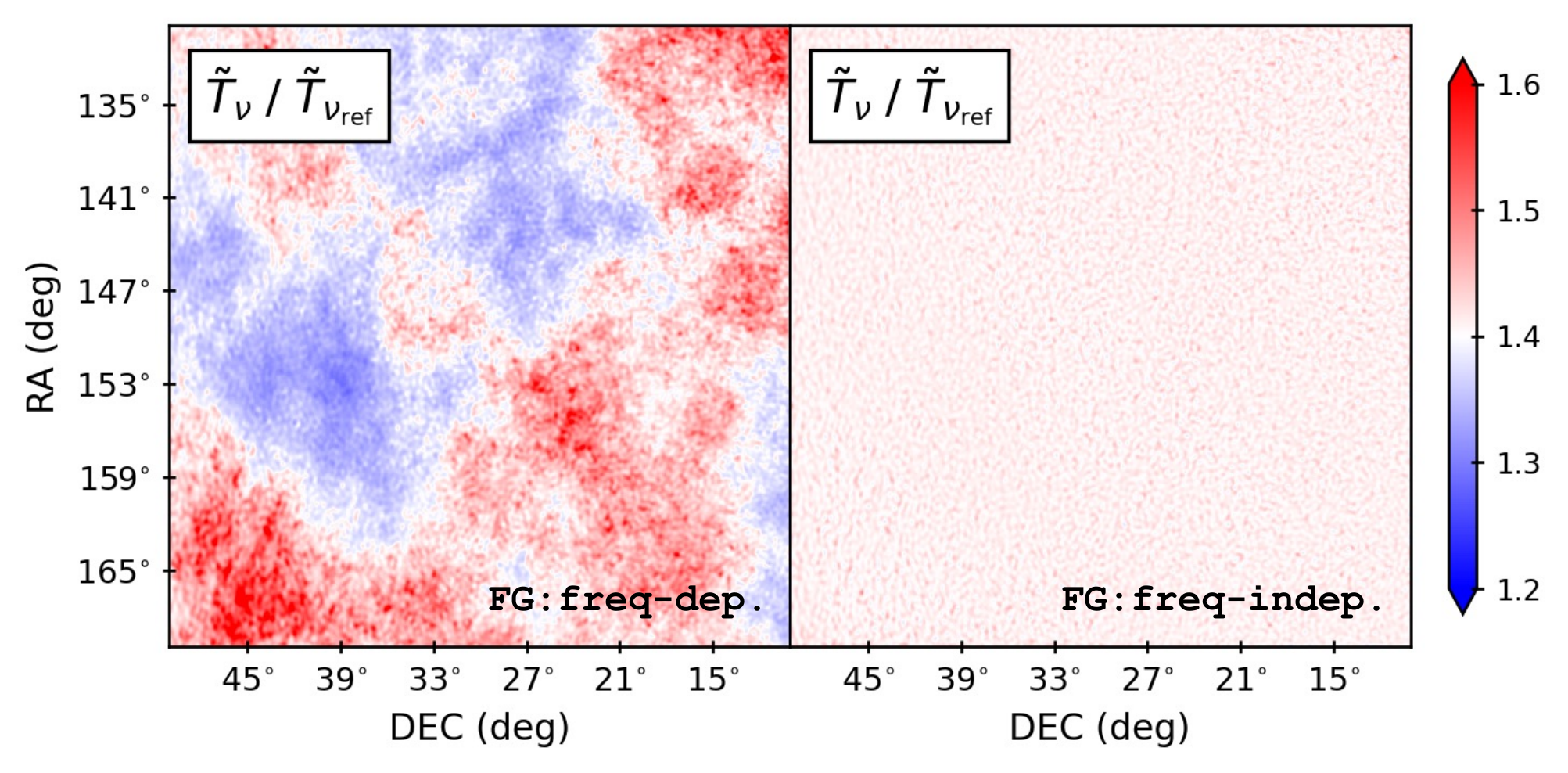}
    \caption{The temperature ratio $(\widetilde{T}_{\nu}/\widetilde{T}_{\nuref})$ between  $800.2 {\rm MHz}$ and the reference frequency $\nuref = 700.0 {\rm MHz}$.  
    To assess the impact of the intrinsic frequency variations in the sky, as described in Eq. \ref{eqn:F_T_filter}, we consider a simplified foreground model, denoted as `$\mathtt{FG:freq-indep}$', in which the spatial patterns at a reference frequency $\nuref$ are held constant, while the average temperature corresponds to the actual temperature at each frequency. The standard, more complex foreground model is referred to as `$\mathtt{FG:freq-depend.}$'. The left panel displays the temperature ratio map using the original sky model, and the right panel illustrates the same for the simplified `$\mathtt{FG:freq-indep}$' model. Notably, in the right panel, the observed noise originates from the independently added HI signal. }
    \label{fig:ratio-Tnu diffFG-sameFG}
\end{figure}

\subsection{Sky Model}
\label{subsec:skymodel}

\begin{figure}
\centering
\includegraphics[width=0.5\textwidth]{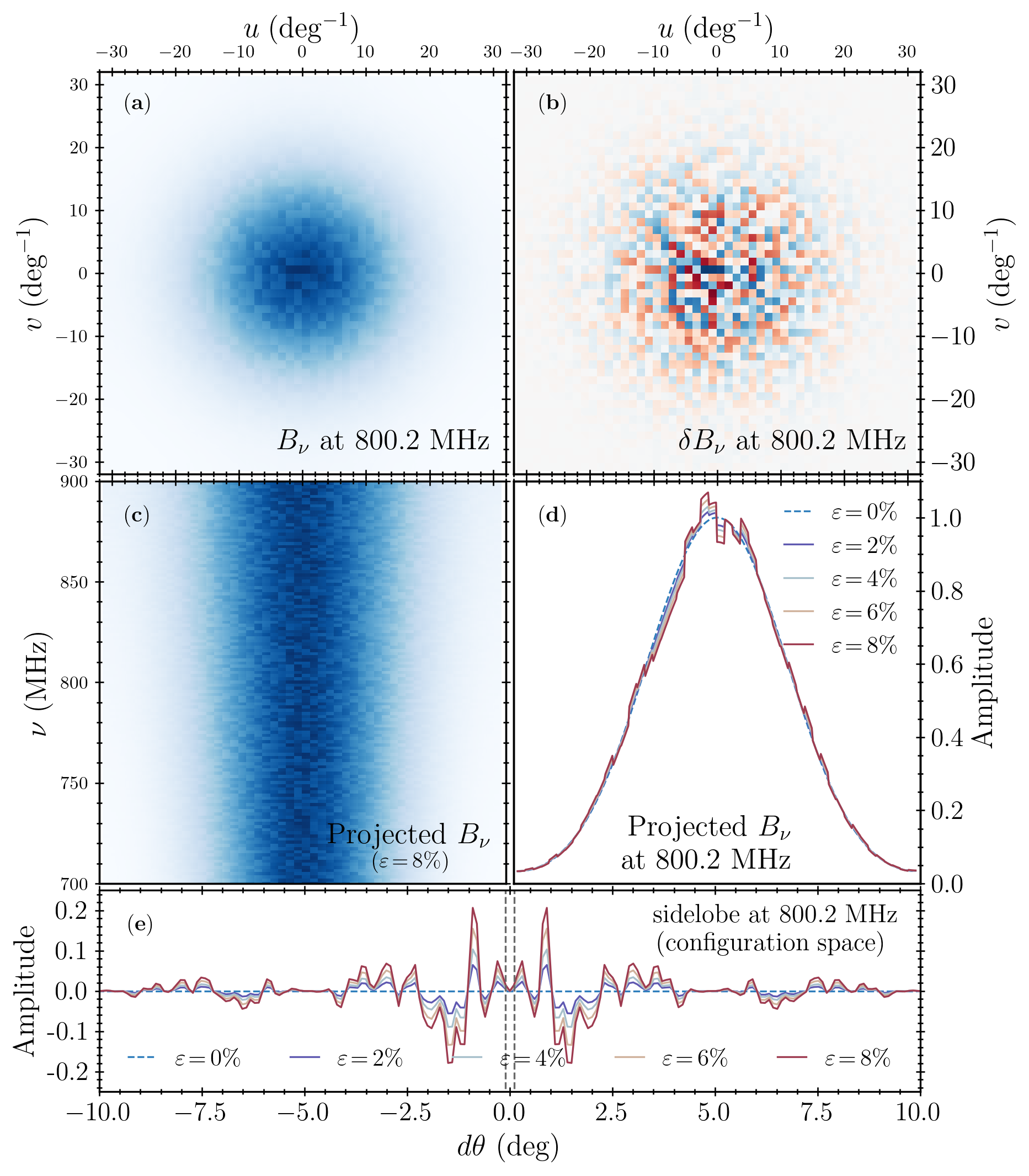}
\caption{Demonstration of our noisy beam Model-I, i.e. the three-dimensional random noise model, for a $100$-meter single-dish telescope:
Panel (a) displays the noisy beam, $B_{\nu}(\vu)$ (Eq.~\ref{eq:1st beam model}), at $800.2,{\rm MHz}$ in Fourier space.
Panel (b) showcases the beam fluctuation $\delta B_{\nu}(\vu)$, with a color scheme representing positive values in red and negative values in blue.
Subplot (c) illustrates the frequency variation of the one-dimensional beam, spanning from $700$ to $900 {\rm MHz}$.
Plot (d) presents the one-dimensional beams at $800.2~ {\rm MHz}$ for five different values of the noise amplitude $\varepsilon=(0\%, 2\%, 4\%, 6\%, 8\%)$.
Panel (e) visualizes the side-lobes in configuration space generated by inverse-Fourier transform of the beam fluctuation $\delta B_{\nu}(\vu)$. The normalization is applied such that the fluctuation is zero at the center. The dashed lines indicate the size of the Gaussian main beam $\theta_{\rm\tiny FWHM}$ at $800.2~{\rm MHz}$.}
\label{fig:Beam_Fluc_with_sidelobe}
\end{figure}

For simplicity, the sky model for our simulation is generated using the publicly available $\mathtt{CRIME}$ code \cite{alonso2014fast} \footnote{\url{http://intensitymapping.physics.ox.ac.uk/CRIME.html}},  
which includes five distinct components:
(a) HI signal,
(b) Galactic synchrotron emission,
(c) Galactic free-free emission,
(d) Extra-galactic free-free emission,
(e) Extra-galactic point sources emission.
As demonstrated in \cite{alonso2014fast}, the HI signal here is derived from the lognormal model \citep{Coles1991MNRAS} of matter fluctuation $\delta(\vx)$, which is then transformed into a map of 21cm brightness temperature fluctuations
\begin{equation}
    T^{\rm HI}(\los, z) =(0.19055\, \mathrm{K})\frac{ h\,\Omega_{b}(1+z)^{2} x_{\rm HI}(z) }{ \sqrt{\Omega_{m}(1+z)^{3}+\Omega_{\Lambda} }}
    (1+\delta_{\rm HI}), 
\end{equation}
where $\delta_{\rm HI}$ represents the HI overdensity, smoothed over the volume element defined by $\delta\nu$ and $\delta\Omega$. Here the HI fraction $x_{\rm HI}(z)=0.008(1+z)$ is assumed to be fixed. Although a more complex modeling approach might be beneficial, it is not critical for the purposes of our study. In this paper, we consider the frequency range between  $700\!-\!900\, \rm MHz$ corresponding to the redshift range $0.58\!<\!z\!<\!1.03$ in our choice of cosmology. 

In terms of foreground modeling, $\mathtt{CRIME}$ models the large-scale Galactic synchrotron radiation (b) by extrapolating the $408\,\rm MHz$ Haslam map~\citep{haslam1982408}. This extrapolation utilizes a direction-dependent spectral index derived from the Planck Sky model~\citep{delabrouille2013pre}, formulated as:
\begin{equation}
    T^{\rm sync}(\nu, \mathbf{\hat{n}}) = T_{\rm Haslam}(\mathbf{\hat{n}})\left (\frac{\nu_{H}}{\nu} \right )^{\beta(\mathbf{\hat{n}})},
\end{equation}
where $\beta(\los)$ is the synchrotron spectral index, varying with the line-of-sight (LoS) direction $\los$. To capture structures on scales smaller than Haslam resolution, $\mathtt{CRIME}$ incorporates a constrained Gaussian realization with parameters from \citep{santos2005multifrequency,alonso2014fast}. Similarly, for the Galactic free-free (c) and Extra-galactic point sources (e) emissions, $\mathtt{CRIME}$ also uses Gaussian realizations based on specific angular power spectra \citep{alonso2014fast}.

From the full-sky simulated maps, we then cutout a mock data cube measuring $40^{\circ} \times 40^{\circ}$ in angular direction and spanning a frequency range of $700 - 900{\rm MHz}$.  Due to computational constraints, the angular resolution of the maps is set at $3.44~\rm arcmin$, corresponding to $\mathtt{nside}=1024$ in the $\mathtt{healpix}$ format.

As illustrated in Figure (\ref{fig:The_selected_sky-region}), we have chosen a relatively clean region with right ascension $130^{\circ} \le {\rm RA} \le 170^{\circ}$ and declination $10^{\circ} \le {\rm DEC} \le 50^{\circ}$. In the frequency domain, we have divided the data into 128 frequency channels spanning the range from $700\,{\rm MHz}$ to $900\,{\rm MHz}$. Subsequently, we proceed to grid this dataset into a three-dimensional cube with dimensions $(N_{\nu}, N_{\rm RA}, N_{\rm DEC})=(128,400,400)$. This raw data cube will then be convolved with beam functions, as detailed in the following subsection, Sec.~\ref{subsec:simbeam}.

\begin{figure}
    \centering
    \includegraphics[width=0.48\textwidth]{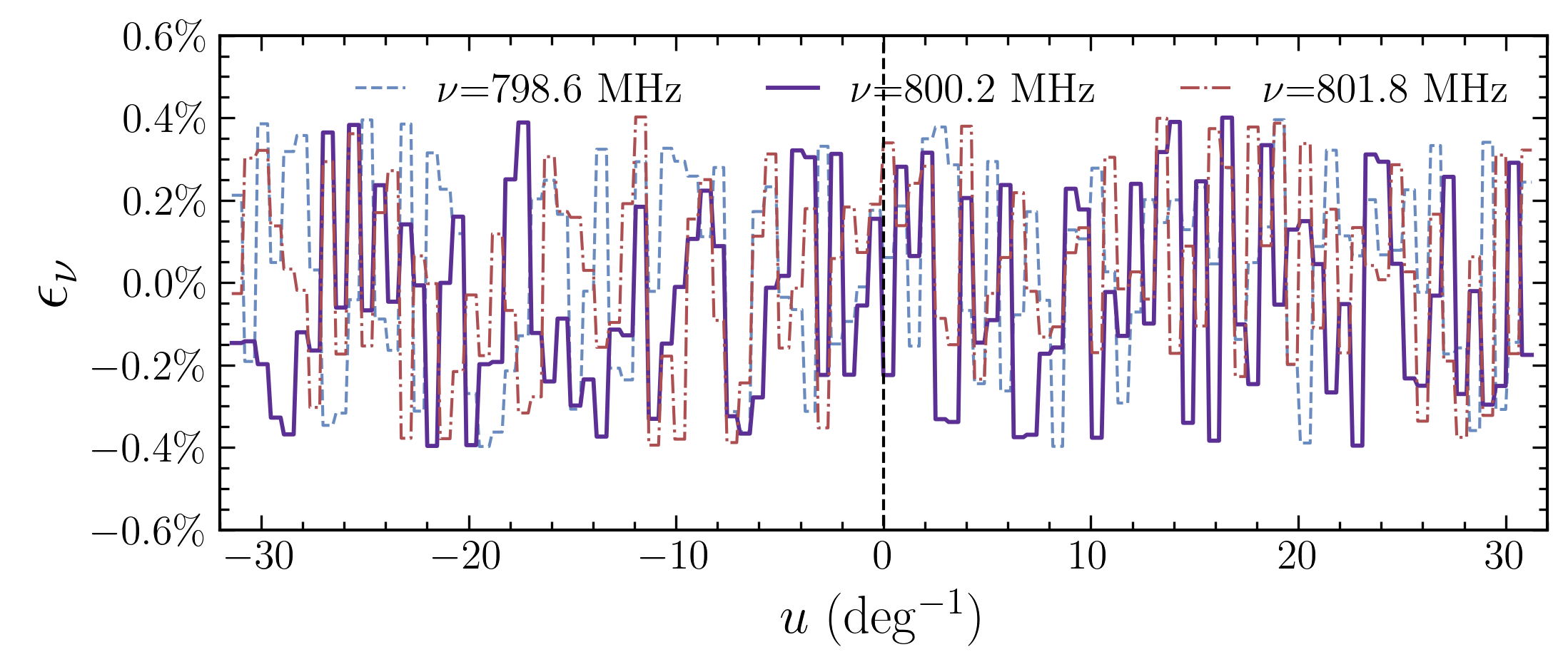}
    \includegraphics[width=0.48\textwidth]{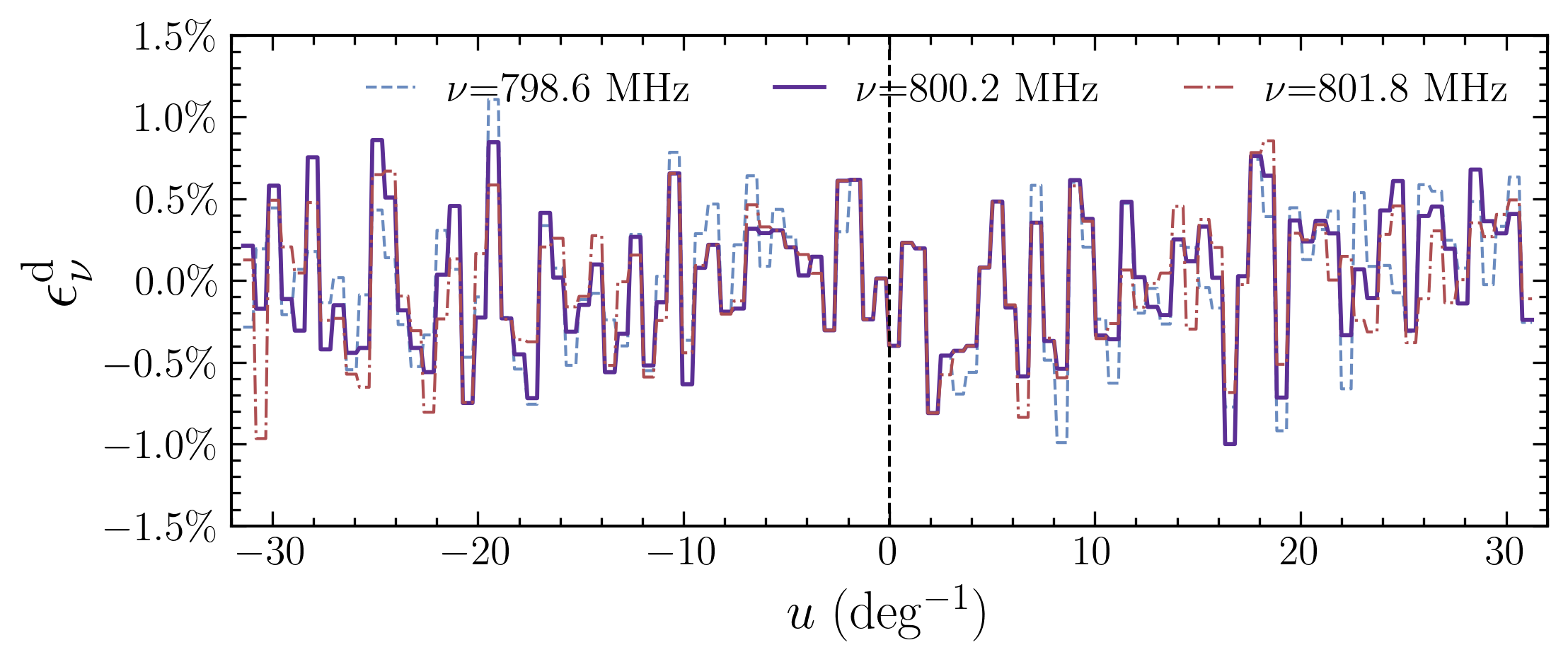}
    \caption{
    The one-dimensional beam fluctuation in Fourier space, with different colored lines indicating three distinct frequencies around $800\,{\rm MHz}$. 
    The upper panel illustrates the three-dimensional random noise model, i.e., the Model-I, \rev{with $\varepsilon=0.4\%$.} As observed, in this case, fluctuations are entirely random across frequencies, potentially leading to a more pronounced impact on the foreground removal.
    The lower panel presents the same plot for Model-II \rev{with $\varepsilon=1.6\%$}, where the noise originates physically within the telescope's spatial configuration. From the plot, it is evident that the noise appears similar at low $u$ values for nearby frequencies and only starts to diverge at larger $u$. 
    }
    \label{fig: Beam_Fluctuation}
\end{figure}

Furthermore, to evaluate the impact of the intrinsic frequency variations in the sky (as described in Eq. \ref{eqn:F_T_filter}), we consider a simplified foreground model in which the spatial patterns at a reference frequency $\nuref$ are held fixed, while the overall amplitudes are adjusted according to the respective frequency. In the following, this model is labeled to as ``$\mathtt{FG:freq-indep}$'', whereas the original foreground model is denoted as ``$\mathtt{FG:freq-depend.}$''. In Figure (\ref{fig:ratio-Tnu diffFG-sameFG}), we display the map of temperature ratio between two frequencies: $\widetilde{T}_{\nu}(\los)$ at $800.2 {\rm MHz}$ and the reference frequency $\nuref = 900 {\rm MHz}$. The left panel displays the original $\mathtt{freq-depend.}$ sky map, while the right panel showcases the simplified $\mathtt{freq-indep.}$ model. In the simplified model, we observe some noise around the value of $1.4$, which is attributed to the independently added HI signal.

\subsection{Beam Model}
\label{subsec:simbeam} 

In this subsection, we introduce the noisy beam model utilized in this study. 
For simplicity, we consider a single dish telescope with a Gaussian model as the unperturbed beam
\begin{eqnarray}
\label{eq: Gaussian beam}
    B^G(\theta) = \exp\left( -\frac{\theta^2}{2\sigma^2} \right)
\end{eqnarray}
with $\sigma = \theta_{\rm \tiny FWHM} /(2\sqrt{2\ln 2}) $. The full-width at half-maximum (FWHM) $\theta_{\rm \tiny FWHM}$ is determined by 
\begin{equation}
\label{eq:theta_FWHM}
    \theta_{\rm \tiny FWHM} = \frac{c}{\nu D},
\end{equation}
where $D$ is the diameter of the telescope dish. In many instances, this model offers a sufficiently accurate approximation. \rev{For example, as demonstrated in \cite{Liao2016ApJ}, it provides a robust first-order approximation for both the intensity and polarization of the Green Bank Telescope. }

\begin{figure}
    \centering
    \includegraphics[width=0.48\textwidth]{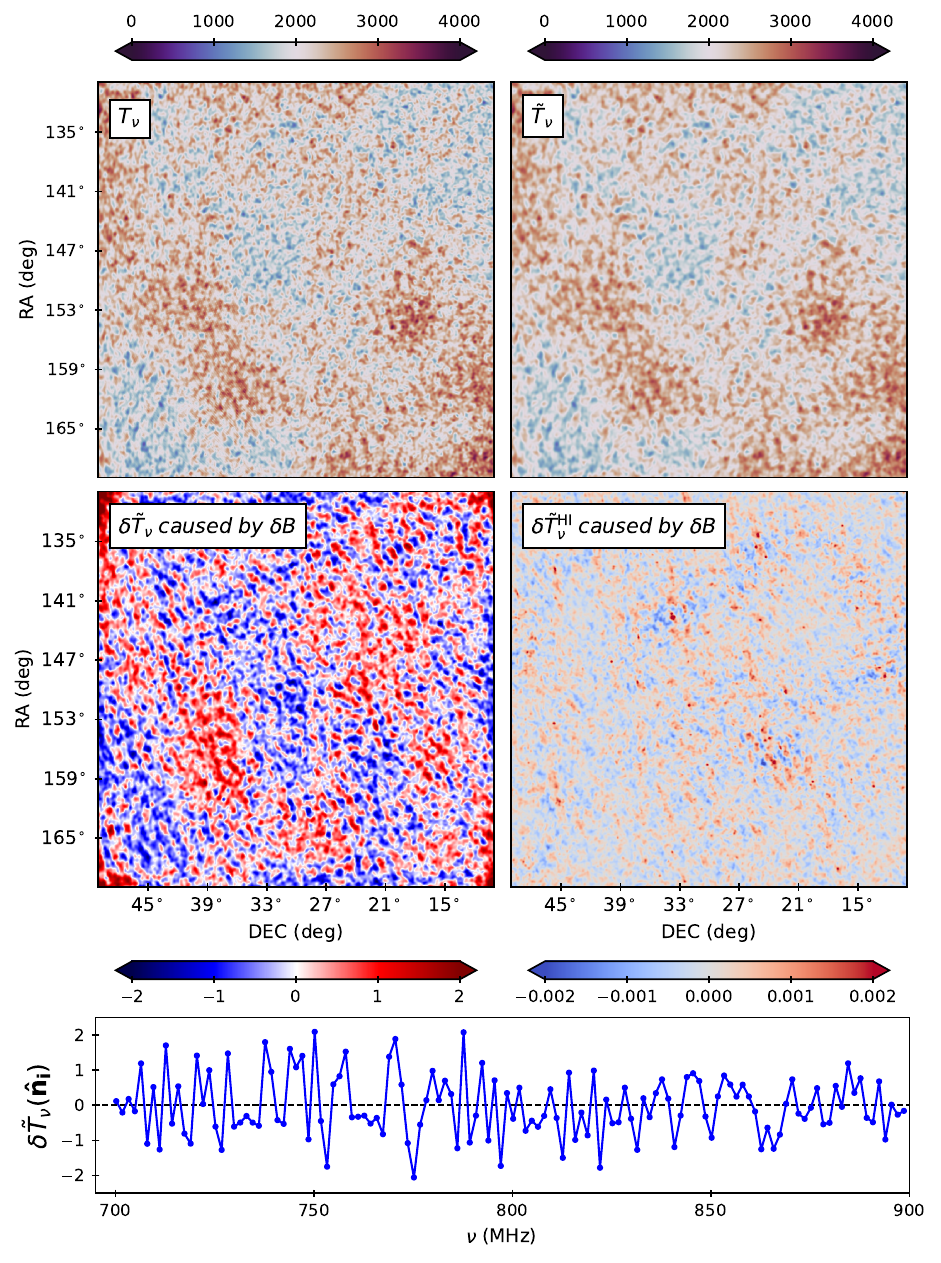}
    \caption{Comparison between the raw sky map ({\it upper-left} panel, with unit in $\rm mK$) and its convolution ({\it upper-right} panel) with the noisy beam \rev{Model-I ($\varepsilon=0.4\%$)} $B$  at $800.2 ~{\rm MHz}$.
    The {\it middle-left} panel demonstrates the residual map $\delta\widetilde{T}_{\nu}$ caused by the beam noise $\delta B_{\nu}$ at $\nu=800.2 ~ {\rm MHz}$. The {\it middle-right} panel displays the residual HI map $\delta\widetilde{T}^{\rm HI}_{\nu}$ at the same frequency and beam noise level. 
    The {\it bottom} panel shows the frequency variation of the temperature residual $\delta\widetilde{T}_{\nu}(\los_{\rm i})$ along the frequency.
    }
    \label{fig:T_raw-beam}
\end{figure}

As demonstrated in section \ref{sec:method}, our method is implemented in Fourier space. In this domain, the Gaussian beam is represented as 
\begin{eqnarray}
B^G_{\nu}(\vu)&=&\exp\left(-\frac{\sigma(\nu)^2 u^2}{2}\right). 
\end{eqnarray}
In the following, we focus on a single dish telescope resembling the Green Bank Telescope (GBT) with a $100$-meter diameter, corresponding to the angular resolution $\theta_{\rm \tiny FWHM}$ that varies from $14$ to $18~{\rm arcmin}$ across the frequency range $\nu=700\!-\!900~{\rm MHz}$.

In this paper we consider two simple models of beam fluctuation.  The first model (Model-I) treats the beam fluctuation as a three-dimensional random field in the $\vu-\nu$ space, 
\begin{eqnarray}
\label{eq:1st beam model}
    B_{\nu}(\vu)&=&B^{G}_{\nu}(\vu)+\delta B_{\nu}(\vu) \\ \nonumber
    &=& B^{G}_{\nu}(\vu) (1+\epsilon_{\nu}(\vu)),
\end{eqnarray}
where $\delta B_{\nu}(\vu)$ is the beam noise, and $\epsilon_{\nu}(\vu)$ is a zero mean random field   in $\vu-\nu$ space, defined as
\begin{eqnarray}
\label{eq:beam01-fluc}
    \epsilon_{\nu}(\vu) = \varepsilon~{\rm rand}(\vu, \nu).
\end{eqnarray}
Here, $\varepsilon$ is the amplitude coefficient of the noise, and ${\rm rand}$ is a uniformly distributed random variable within the range of $-1$ to $1$. Given the entirely random nature of the noises across frequencies, this model represents a pessimistic scenario regarding potential beam noise.

In our second model of beam noise (Model-II), we consider a more realistic scenario where the noise originates within the telescope's spatial configuration. This noise manifests as a two-dimensional random field
\begin{eqnarray}
\label{eq:bfluc-tspace}
\epsilon^d(\mathbf d)=\varepsilon~{\rm rand}(\mathbf d)
\end{eqnarray}
where $\mathbf{d}$ represents the distance within the telescope's frame. This noise field is subsequently rescaled and projected onto the $\vu$ space at various frequencies using the relation $\vu = {\mathbf d}/\lambda$
\begin{eqnarray}
\label{eq:beam02-fluc}
    \epsilon^d_{\nu}(\vu) = \epsilon^d (\mathbf{d}/\lambda)
\end{eqnarray}

\begin{figure}
    \centering
    \includegraphics[width=0.46\textwidth]{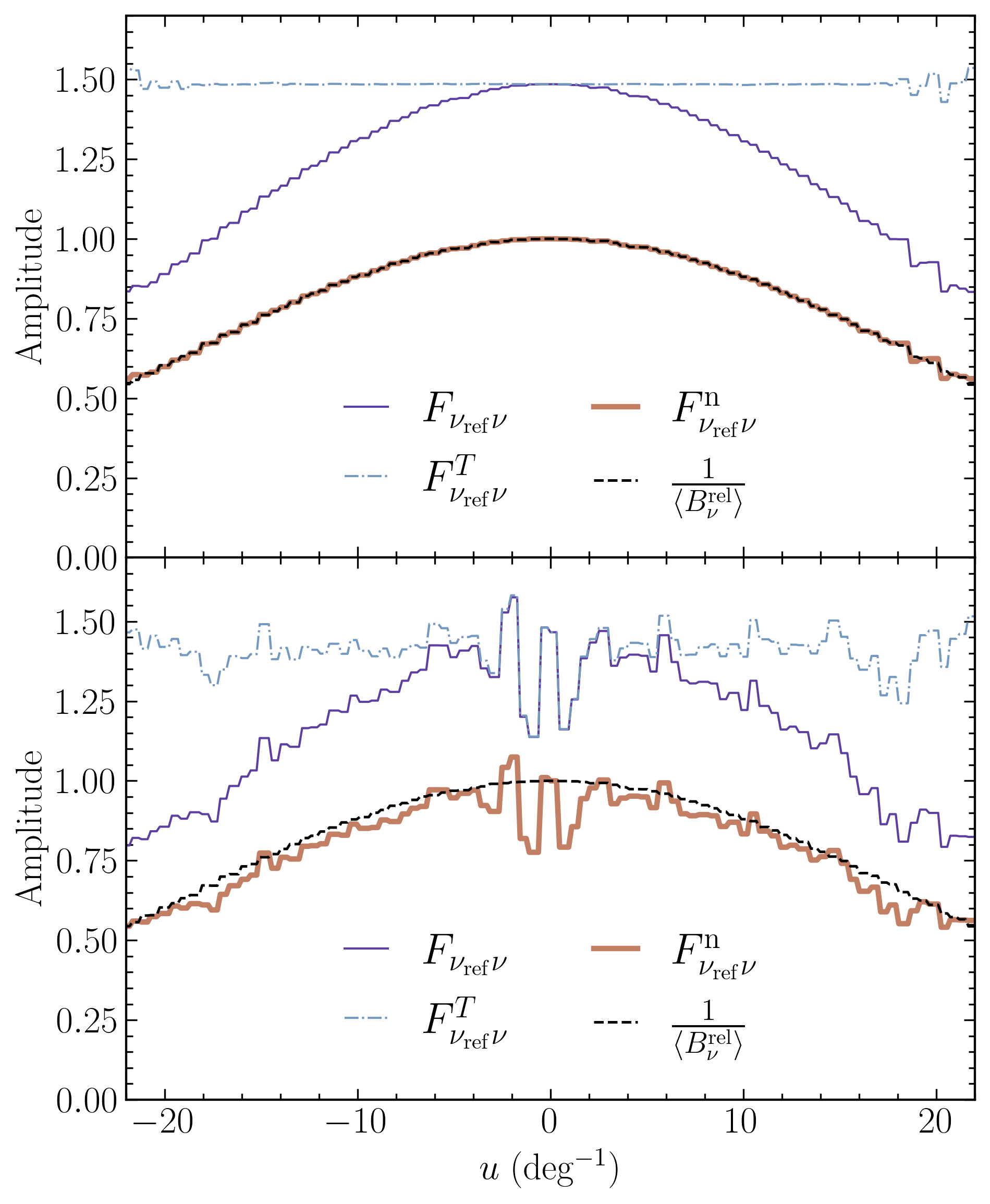}
    \caption{Demonstration of the one-dimensional filter $F_{\nuref\nu}$ defined in Eq.~\ref{eqn:wienerfilter} ({\it purple solid line}) at frequency $\nu$, and the true noisy beam ({\it black dotted line}) injected in the simulation. Here,  $\nuref$ is the referenced frequency chosen to be $700.0~{\rm MHz}$. 
    \rev{Here, we illustrate only Beam Model-I, as the results are very similar to those for Model-II.} 
    To evaluate the impact from intrinsic foreground variation, we demonstrate the filter for the simplified `$\mathtt{freq-indep.}$' noise model in the {\it upper panel} and the realistic `$\mathtt{freq-depend.}$' model in the {\it lower panel}. 
    For better understanding, we also decompose the filter into separate contributions, including the intrinsic temperature factor $F^{T}_{\nuref\nu}$ defined in Eq.~\ref{eqn:F_T_filter} ({\it blue dash-dotted lines}), and the relative beam $B^{\rm rel}_{\nu}$ ({\it black dashed line}). $F^{\rm n}_{\nuref\nu}$ is the normalized filter $F_{\nuref\nu}$ ({\it red solid line}), given by $F^{\rm n}_{\nuref\nu}(\vu) = F_{\nuref\nu}(\vu) / F_{\nuref\nu}(\vu\!=\!0)$.
    }
    \label{fig:Filter diffFG}
\end{figure}

In Figure (\ref{fig:Beam_Fluc_with_sidelobe}), present our noisy beam model for a $100$ meter single-dish telescope, showcasing in both Fourier space (panels a-d) and configuration space (panel e). Panels (a) and (b) illustrate the noisy beam $B_{\nu}(\vu)$ at $800.2,{\rm MHz}$ and its noise component $\delta B_{\nu}(\vu)$, respectively. The resolution of $\delta B_{\nu}(\vu)$ in Fourier space is about $1.3~{\deg}^{-1}$, as shown in panel (b), with positive values indicated in red and negative values in blue. Model-I is characterized by completely random fluctuations in the frequency direction. This is exemplified in panel (c), which depicts a slice of the one-dimensional beam as a function of frequency. Panel (d) displays the one-dimensional beam at a specific frequency ($800.2 {\rm MHz}$) under various noise levels $\varepsilon$, while keeping the underlying random field constant. Here, $\varepsilon = 2\%$ suggests that the beam fluctuates within a range of $-0.02$ to $0.02$. Finally, panel (e) illustrates the noisy side-lobes in configuration space, resulting from the beam noise $\delta B$. The normalization chosen here ensures $B(\theta=0)=1$ at the center. We notice that these side-lobes in configuration space, emerged from simple Fourier space random noise, do resemble features in actual telescope beams, highlighting the practicality of our beam noise model.

\begin{figure}
    \centering
    \includegraphics[width=0.46\textwidth]{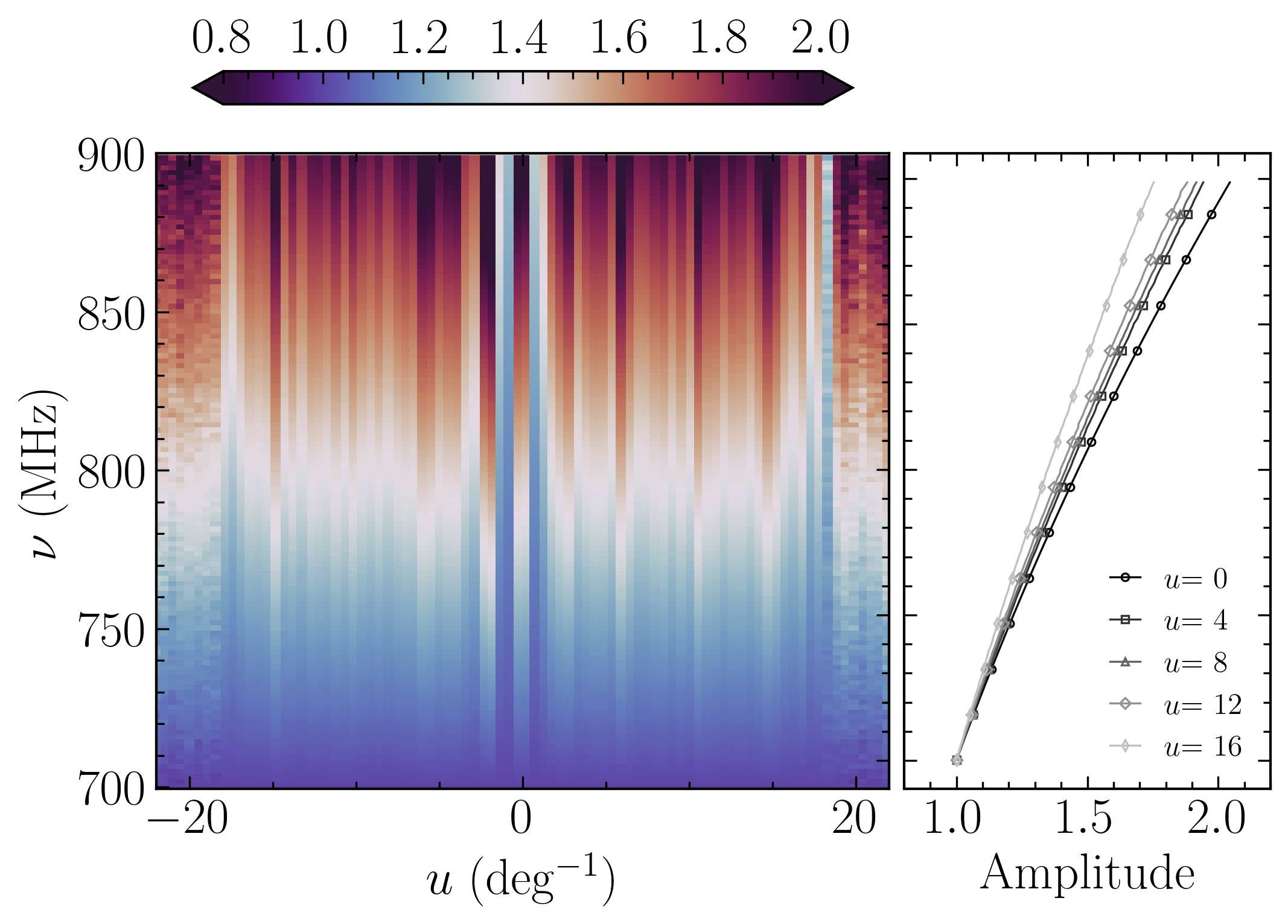}
    \caption{ 
    Frequency variation of the intrinsic temperature ratio $F^{T}_{\nuref\nu}$ for the `$\mathtt{freq-depend}$' foreground model. With the reference frequency set at $\nuref=700 {\rm MHz}$, the amplitude of the factor $F^T$ gradually increases. Although this factor exhibits significant spatial variation, its frequency dependence remains quite smooth ({\it right panel}). This characteristic results in a comparatively lesser impact on the final foreground removal process than the random beam noise we introduced. }
    \label{fig:F-T-diffFG}
\end{figure}

In Figure (\ref{fig: Beam_Fluctuation}), we showcase the one-dimensional beam fluctuation in Fourier space, with different colors representing three distinct frequency channels around $800 \, {\rm MHz}$. The upper panel illustrates our `pessimistic' Model-I with $\varepsilon=0.4\%$, which is the three-dimensional random noise model, described by Eq. (\ref{eq:beam01-fluc}). In this model, the beam fluctuations are completely independent across frequencies, indicating a potentially larger impact on foreground removal. The lower panel shows a similar plot for Model-II with $\varepsilon=1.6\%$, where the noise resides in the telescope's spatial domain. In this model, the frequency-dependence of the noise arises from the chromatic relationship described in Eq. (\ref{eq:beam02-fluc}). From the plot, it is clear that the noise profiles appear similar at lower $k$ values for nearby frequencies, with deviations only becoming noticeable at higher $k$ values.
\rev{Our choice of $\varepsilon$ here is consistent with real-world beam fluctuations. For instance, \cite{Liao_2016} measured the beam pattern of the Green Bank Telescope (GBT) using astrophysical sources such as quasars and pulsars, characterized by two-dimensional Gauss-Hermite coefficients. 
Using their beam model, we estimate that the root-mean-square (RMS) of the beam fluctuation along the frequency direction could reach as high as $0.8\%$. }

\subsection{Mock Observation}
\label{subsec:mock obsmap} 

After convolving our simulated datacube with the noisy beam model, we derive the mock observation data $\widetilde{T}_{\nu}(\vu)$ in Fourier space
\begin{align}
    \widetilde{T}_{\nu}(\vu)  &= B_{\nu}(\vu)\, T_{\nu}(\vu)
\end{align}
as well as the corresponding map in real space $\widetilde{T}_{\nu}(\los)$. 

Figure (\ref{fig:T_raw-beam}) offers a comparative view of the mock observation data against the raw temperature map. The upper panels show the raw sky map ({\it upper-left}), denoted as $T_{\nu}$, alongside its convolution with the noisy telescope beam $\widetilde{T}_{\nu}$ ({\it upper-right}), under beam Model-I (Eq.~\ref{eq:1st beam model}), at a frequency of $800.2 {\rm MHz}$. In the middle-left panel, we illustrate the difference between these two plots, effectively highlighting the temperature variations caused by the beam noise of the telescope. Similarly, the middle-right panel displays the HI variation caused by the beam noise. Finally, the bottom part of the figure demonstrates the temperature fluctuation, $\delta \widetilde{T}$, along the frequency axis. It is evident from this display that such frequency variations will significantly challenge the effectiveness of foreground subtraction methods.

\begin{figure*}
\centering
\includegraphics[width=0.9\textwidth]{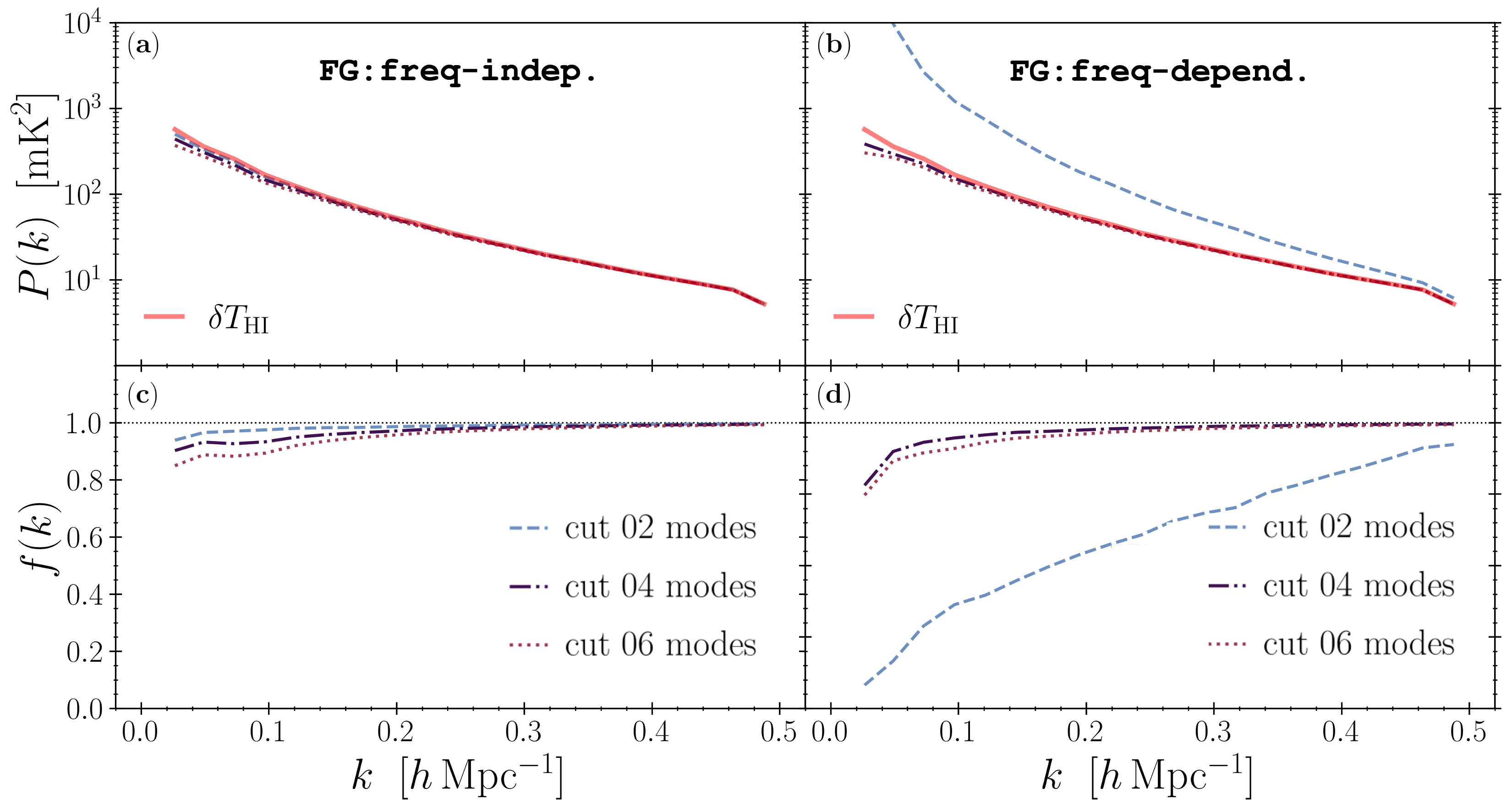}
\caption{Auto powerspectrum ({\it upper} panels) and cross-correlation ratio ({\it lower} panels) after SVD foreground removal, under the assumption that the relative beam $f_{\nuref\nu}(\vu)=B_{\nuref}/B_{\nu}$ is precisely known, here $B_{\nu}=B^{G}_{\nu}+\delta B_{\nu}$. The {\it left} panels shows the case of the `$\mathtt{freq\!-\!indep}$' foreground model, the right column corresponds the case of the `$\mathtt{freq\!-\!dep}$' foreground.}
\label{fig:auto-cross-trueB}
\end{figure*}

\section{Results}
\label{sec:results}
Before proceeding to foreground subtraction and statistical analysis, our CBC method constructs a frequency-dependent filter $F_{\nuref\nu}(\vu)$. As detailed in (\ref{sec:method}), this filter is generated through the cross-correlation of maps in the $\vu$ space, comparing each frequency $\nu$ with a reference frequency $\nuref$. This section aims to demonstrate the effectiveness of this approach. We will first discuss the calibration of the beam, followed by showcasing the results achieved through foreground removal. 

\subsection{Beam Calibration}
\label{sec:beam_calibration}

In an ideal scenario, after applying the filter $F_{\nuref\nu}$, observational maps at different frequencies should all be convolved with the same beam, specifically the one at the reference frequency $B_{\nuref}$. However, in practice, intrinsic temperature variations across frequencies complicate this process, as they introduce an additional temperature factor (Eq. \ref{eqn:Ffilter}). To dissect the separate impacts of these two factors, we also consider a simplified `$\mathtt{freq-indep.}$' foreground model. In this model, the temperature factor $F^T_{\nuref\nu}$ is held constant in the $\vu$ space. Meanwhile, the $\nu$-dependent overall temperature amplitude of each channel is adjusted to aligns with that of the actual foreground model.

In the upper panel of Figure (\ref{fig:Filter diffFG}), we illustrate a one-dimensional slice of the filter $F_{\nuref\nu}$ along with its individual components. Here, the sky temperature ratio $F^{T}_{\nuref\nu}$ remains constant ({\it blue dash-dotted line}) at a value of approximately $1.5$. Consequently, the normalized constructed filter $F^n_{\nuref\nu}$ ({\it red solid line}) closely resembles the relative beam $1/B^{\rm rel}_{\nu}$ ({\it dotted line}) that we introduced in the beam model. This convincingly validates the efficacy of our method. 

In the lower panel of Figure (\ref{fig:Filter diffFG}), we present the filter for the more realistic `$\mathtt{freq-depend.}$' foreground. In contrast to the `$\mathtt{freq-indep.}$' model, the intrinsic variation in sky temperature ({\it blue dash-dotted line}) has a substantial influence on the estimation of the relative beam. 
As depicted, the fluctuations in the temperature ratio are considerably larger than the beam noise we introduced ({\it black dashed line}). As a result, the constructed filter ({\it red solid line}) bears a closer resemblance to the temperature ratio than the relative beam. 
However, this extra factor actually does {\it not} invalidate our method. In fact, as we will demonstrate in the next subsection, the foreground-cleaned map following the aforementioned procedure still exhibits considerable improvements compared to traditional methods. This is primarily because the variation in sky temperature, as shown in Figure (\ref{fig:F-T-diffFG}), changes {\it gradually} along the frequency axis. In contrast, while the beam noise we introduced might appear small, it is either completely random (beam Model-I) or somewhat random (Model-II) along frequency. This more pronounced frequency dependence has a much more impactful influence on the result of foreground removal.

\begin{figure*}
\centering
\includegraphics[width=0.9\textwidth]{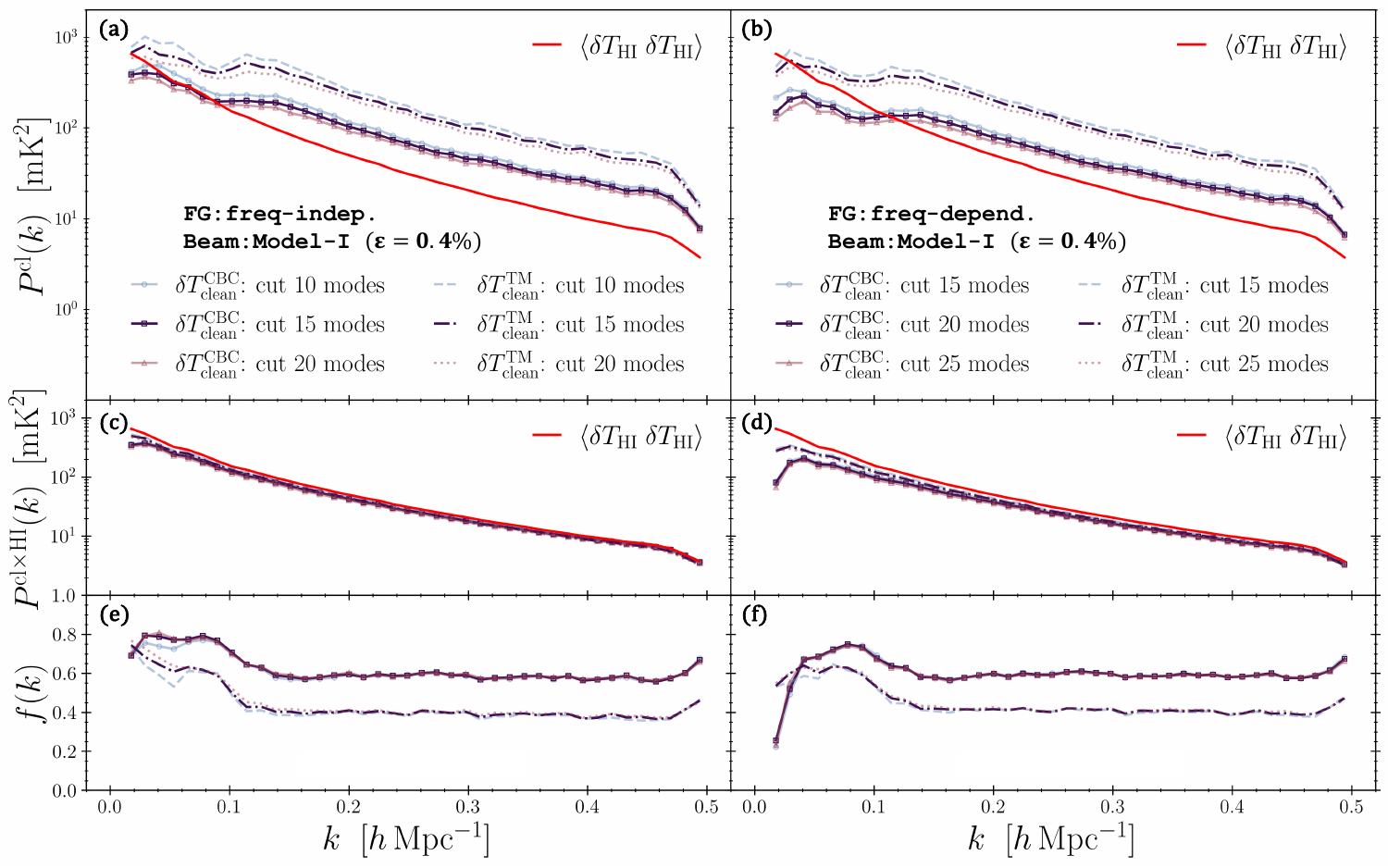}
\caption{Auto- and cross-powerspectrum of SVD foreground removed map \rev{for noisy beam Model-I with $\varepsilon=0.4\%$.} Here, we compare the results between our CBC method ({\it solid} lines) and the traditional approach ({\it dash} and {\it dotted} lines). The {\it upper} panels show the auto-powerspectrum of the cleaned map $\delta T_{\rm clean}$ and {\it lower} panels demonstrate the cross-correlation ratio with injected HI signal. 
The left column shows the case of the `$\mathtt{FG:freq\!-\!indep}$' foreground model, and the right column corresponds to the case of the `$\mathtt{FG:freq\!-\!depend}$' foreground model. As shown, our CBC method can reduce the auto-powerspectrum by a factor of a few and enhance the cross-correlation ratio by about $50\%$.
}
\label{fig:fgremoval_result_Brand}
\end{figure*}

\begin{figure*}
\centering
\includegraphics[width=0.9\textwidth]{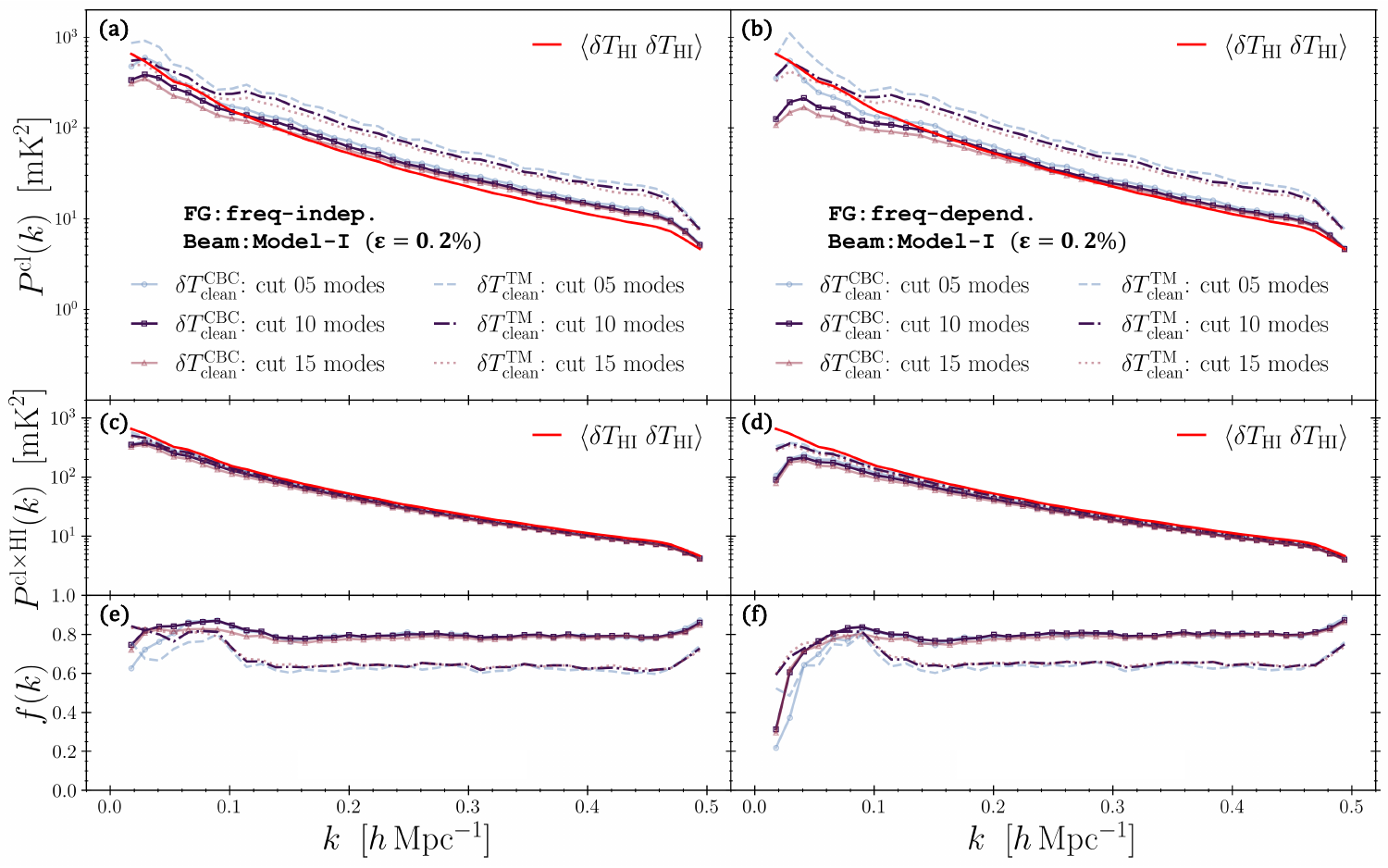}
\caption{\rev{Similar to Figure (\ref{fig:fgremoval_result_Brand}), but for beam Model-I with smaller noise amplitude $\varepsilon=0.2\%$.
Since $\varepsilon$ is only half of the value used in Figure (\ref{fig:fgremoval_result_Brand}), the overall performance is better for both the traditional and CBC method. 
Furthermore, while our CBC method continues to outperform the traditional approach, the correlation ratio for CBC stabilizes around $f\sim 0.8$ compared to $\sim 0.6$ for the traditional method, indicating a reduced improvement factor of about one-third. }
}
\label{fig:fgremoval_result_BM1_eps02}
\end{figure*}

\begin{figure*}
\centering
\includegraphics[width=0.9\textwidth]{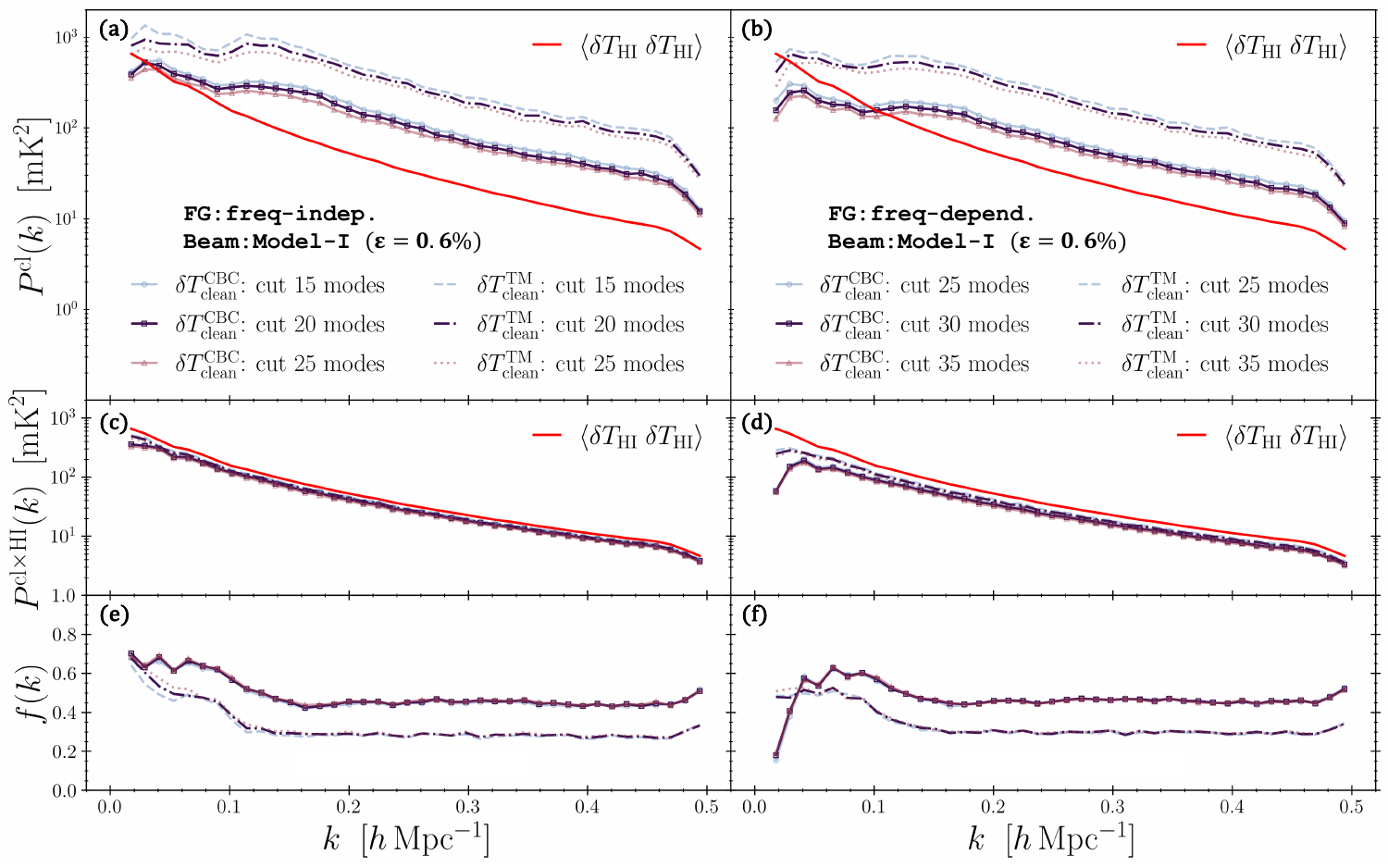}
\caption{ \rev{ Similar to Figure \ref{fig:fgremoval_result_Brand}, but for beam Model-I with higher noise amplitude $\varepsilon=0.6\%$.
As expected, the overall performance has declined for both the traditional and CBC methods. The correlation ratio for CBC is around $f\sim 0.45$ at high $k$ compared to $\sim 0.3$ for the traditional method, demonstrating an improvement ratio of about $50\%$. 
} }
\label{fig:fgremoval_result_BM1_eps06}
\end{figure*}

\begin{figure*}
\centering
\includegraphics[width=0.9\textwidth]{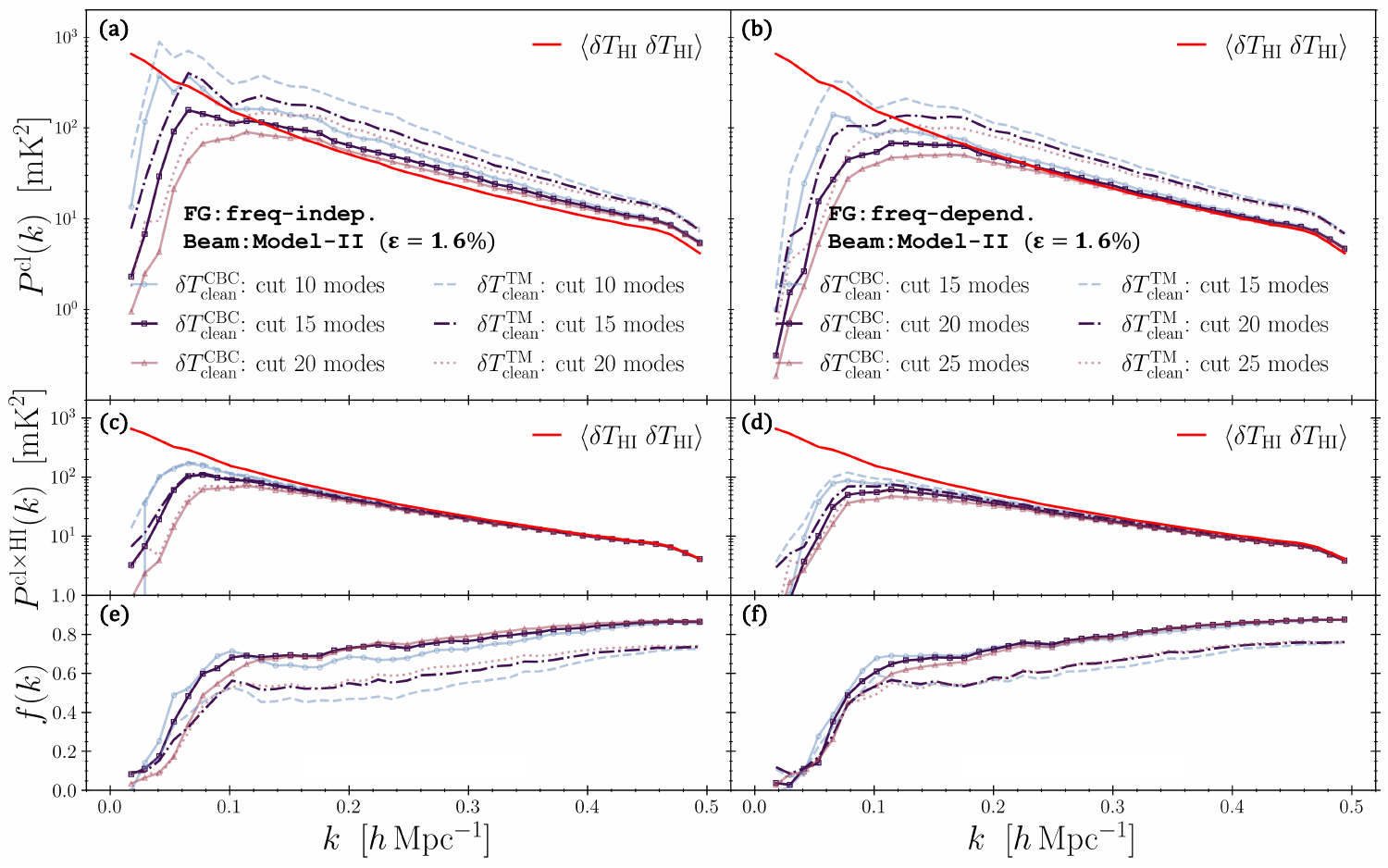}
\caption{\rev{Auto- and cross-powerspectrum of cleaned map for noisy beam Model-II with $\varepsilon=1.6\%$.} Unlike the Model-I, one can observe an increased over-subtraction at larger scales ($k\lesssim 0.15 \mathrm{h/Mpc}$) in both the traditional and CBC methods, regardless of the foreground model considered. However, our CBC method continues to outperform the traditional approach at a similar level seen in Figure (\ref{fig:fgremoval_result_Brand}).}
\label{fig:fgremoval_result_Bmod2}
\end{figure*}

\subsection{Foreground Removal and Statistical Measurements}
\label{sec:res_fgrmv}

After estimating and applying the CBC filter to our mock data, we now proceed with foreground removal and statistical analysis. Specifically, we adopt the blind Principal Component Analysis (PCA) cleaning technique following the methods detailed in \cite{masui2013measurement}  and \cite{switzer2013determination}.
We begin by representing the temperature map as a matrix $\bm{T}$, with dimensions $N_{\nu} \times N_{\theta}$. The PCA modes are then obtained through the decomposition of $\bm{C} =\bm{TT}^{\rm T}/N_{\theta} = \bm{U\Lambda U}^{\rm T}$, where $\bm{\Lambda}$ is a diagonal matrix arranged in descending order. The `cleaned' map, obtained by subtracting the first $\rm m$ PCA modes, is calculated using Eq. (\ref{eqn:pcasub}). 
We continue by calculating the auto-power spectrum of the cleaned map
\begin{eqnarray}
P^{\rm cl}(k)= \left \langle T^{\rm cl}T^{\rm cl} \right \rangle (k).
\end{eqnarray}
Additionally, we also compute the cross-power spectrum with the injected true HI signal. Specifically, our interest lies in the cross-correlation ratio between the two fields, defined as
\begin{eqnarray}
f(k) =\frac{P^{\rm cl\times HI} (k)}
    {\sqrt{\langle T^{\rm HI} T^{\rm HI} \rangle
    \langle T^{\rm cl} T^{\rm cl} \rangle } (k)}
\end{eqnarray}
Here $T^{\rm HI}$ represents the injected HI signal and $P^{\rm cl\times HI}(k)$ is the cross-power spectrum between the signal and the cleaned map. 

As detailed in Section \ref{subsec:skymodel}, our mock observational data encompass the region spanning from $700 {\rm MHz}$ to $900 {\rm MHz}$ in frequency, $130.0^{\circ}$ to $170.0^{\circ}$ in right ascension (RA), and $10.0^{\circ}$ to $50.0^{\circ}$ in declination (DEC). With our chosen cosmological model, this region translate into a comoving box with dimensions of $\left(L_{\nu}, L_{\rm RA}, L_{\rm DEC}\right)=\left(860.8, 1343.7, 1343.7\right)~{\rm  Mpc/h}$.
To assess the effectiveness of our method, we compare the outcomes achieved with the CBC approach to those obtained using a traditional method.
The latter involves convolving data with a common Gaussian beam. This traditional approach, as described in \cite{masui2013measurement}, includes the convolution with an additional kernel, equivalent to Eq. (\ref{eqn:Gaussian_beam_rel-filter}). In this paper, we apply this filter in real space.
Following the discussion in Section \ref{subsec:mock obsmap}, we will conduct a comprehensive comparison that incorporates  two foreground models: the simplified `$\mathtt{freq-indep.}$' model and the realistic `$\mathtt{freq-depend.}$' foregrounds, two beam noise models (Model I and II), and two methods of beam calibration: the traditional method (TM) and our own correlation-based calibration method (CBC).

\begin{figure*}
\centering
\includegraphics[width=0.9\textwidth]{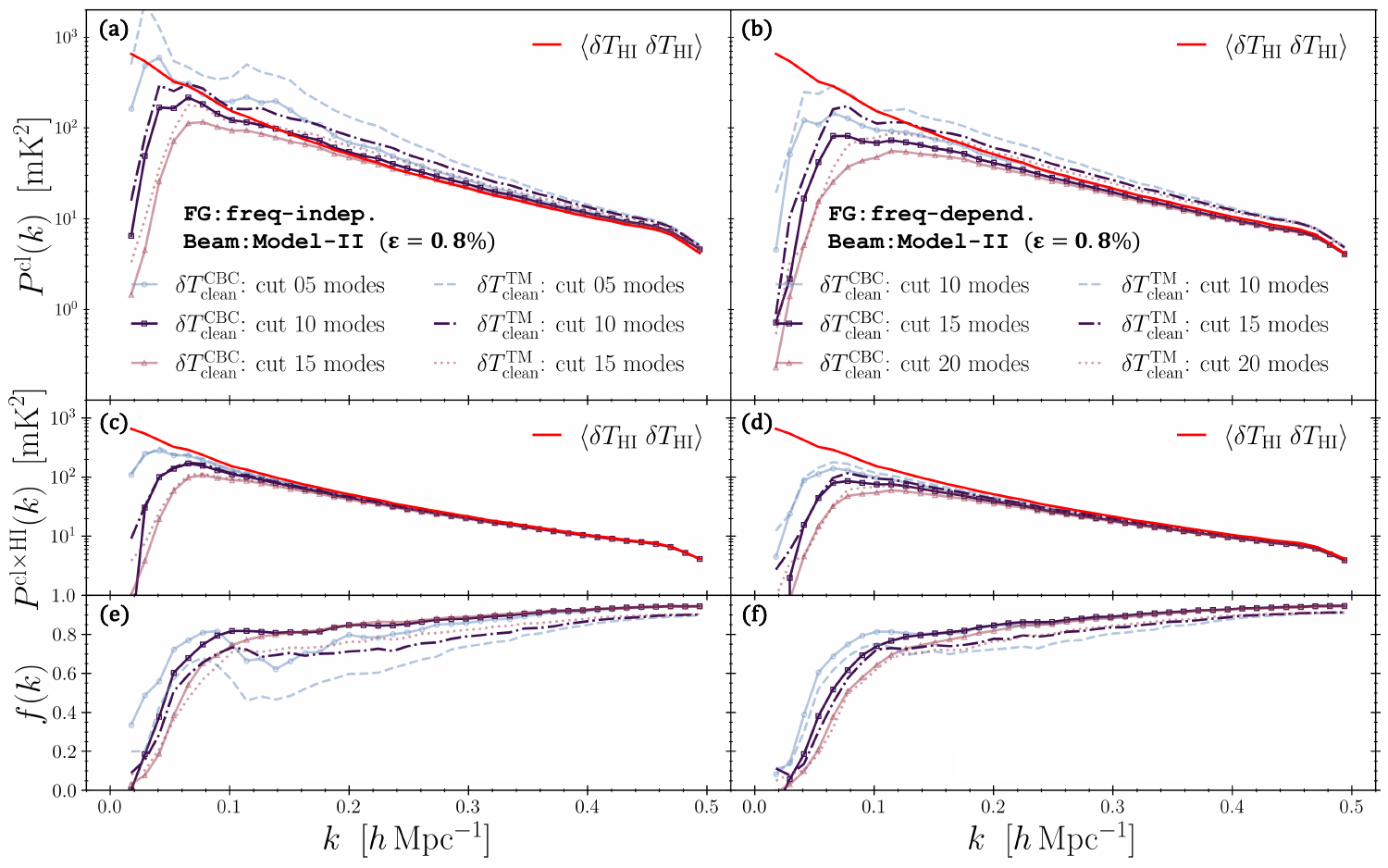}
\caption{\rev{Similar to Figure \ref{fig:fgremoval_result_Bmod2}, but here we assume a smaller $\varepsilon=0.8\%$ for Model-II. As with Model-I, at this lower noise level, both the traditional and CBC methods perform well. The improvement factor for the CBC method is between ten percent and a few percent.  } }
\label{fig:fgremoval_result_Bmod2_eps08}
\end{figure*}

\begin{figure*}
\centering
\includegraphics[width=0.9\textwidth]{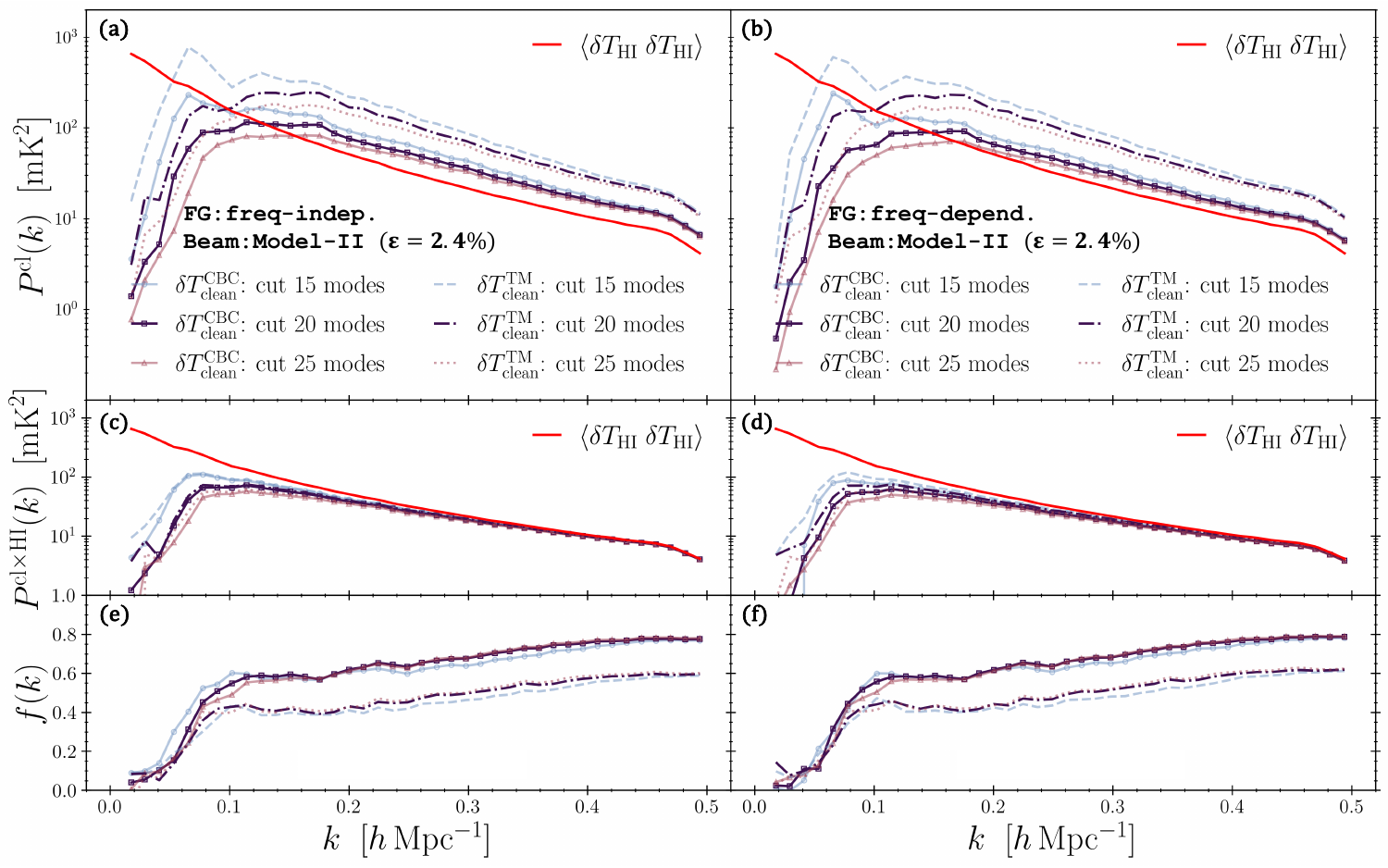}
\caption{\rev{Similar to Figure \ref{fig:fgremoval_result_Bmod2}, assuming $\varepsilon=2.4\%$ for Model-II.  At this noise level, the improvement for the CBC method is approximately one-third.
}}
\label{fig:fgremoval_result_Bmod2_eps24}
\end{figure*}

To establish a baseline result, Figure (\ref{fig:auto-cross-trueB}) presents both the auto power spectrum of the cleaned map ({\it upper} panels) and the cross-correlation ratio with true cosmic signal ({\it lower} panels), assuming that the relative beam $f_{\nuref\nu}(\vu)=B_{\nuref}/B_{\nu}$ is precisely known. As expected, both metrics agree with the expectation of an `ideal' foreground removal, with only a few modes' subtraction resulting in a low auto-power spectrum and a cross-correlation ratio nearing unity. This observation remains consistent across different foreground models and beam calibration methods. However, the results from the simplified `$\mathtt{freq\!-\!indep}$' ({\it left} panels) model demonstrate better performance. Particularly, subtracting as few as $2$ modes ({\it blue-dashed} line) is sufficient to remove this simple foreground contamination. Furthermore, as shown in the cross-correlation,  the more realistic foreground model tends to result in an over-subtraction of the signal at the large-scale ($k \lesssim 0.1 \mathrm{h/Mpc}$). Notice that we did not differentiate between two beam models in this context, as they yield very similar results when the `true knowledge' of beam fluctuations is known. 

We then turn to assessing the efficacy of our CBC method against the traditional approach of the Gaussian common beam regularization. Figure (\ref{fig:fgremoval_result_Brand}) illustrates the result for the three-dimensional random beam model, i.e. Model-I, \rev{with noise amplitude $\varepsilon=0.4\%$}. 
As shown, our CBC method consistently outperforms the traditional approach, regardless of the number of subtracted modes ($N_{\rm sub}$), or the choice of foreground models. For each chosen value of $N_{\rm sub}$, the CBC method significantly reduces the amplitude of the auto-power spectrum by a factor of several and enhances the cross-correlation ratio $f$ by approximately $50\%$. These outcomes suggest that, with such a noisy beam, the CBC method excels in removing foregrounds more effectively and in preserving the HI signals more efficiently.

\rev{For both methods, a sudden shape change appears in the auto-power spectrum around $k\sim 0.1 {\rm h/Mpc}$, resulting in a bump in the cross-correlation ratio $f$. This signature is the residual foreground caused by the mixing of beam noise with the intrinsic structure in the foreground. Such foreground structure can be partially observed in Figure  \ref{fig:Filter diffFG}. }
In the traditional method, we observe that the cross-correlation ratio $f$ undergoes slightly more degradation than in the CBC method, especially at larger scales ($ k\lesssim 0.5 \mathrm{h/Mpc}$). Additionally, the performance impact when using the realistic `$\mathtt{freq\!-\!dep}$' foreground model is surprisingly insignificant compared to the simpler `$\mathtt{freq\!-\!indp}$' model. This can be partly attributed to the slow frequency variation of the temperature ratio $F^T_{\nuref\nu}=T_{\nu}/T_{\nuref}$. Despite the significant spatial variation of $F^T_{\nuref\nu}$ relative to the beam noise, as illustrated in Figure (\ref{fig:Filter diffFG}), its frequency dependence is actually quite smooth. This is evident in Figure (\ref{fig:F-T-diffFG}), where the right panel shows a different but slow frequency change across various spatial $u$ values.

\rev{To examine how the performance of CBC method varies with the noise levels of the beam, we also analyze  the results for two different amplitude, $\varepsilon$, in Model-I.  Figure \ref{fig:fgremoval_result_BM1_eps02} and \ref{fig:fgremoval_result_BM1_eps06} illustrate the outcomes for $\varepsilon=0.2\%$ and $\varepsilon=0.6\%$, respectively.  For smaller beam noise ($\varepsilon=0.2\%$), we are able to subtract fewer modes to achieve a similar level of the cross-power spectrum, and the auto-power spectrum also approaches more closely to the injected signal for both the traditional and CBC methods. Although the CBC method continues to outperform the traditional approach, the correlation ratio for the CBC method stabilizes around $f\sim 0.8$ at large $k$, compared to $f \sim 0.6$ for the traditional method. This indicates a reduced improvement factor of about one-third. 
In contrast, with larger noise ($\varepsilon=0.6\%$),  greater residual foregrounds lead to a higher auto-power spectrum and a lower cross-correlation ratio for both methods. Regardless, the CBC method still improves the cross-correlation ratio by approximately $50\%$, which is a similar level of improvement compared to our fiducial model with $\varepsilon=0.4\%$.
}

Figure \ref{fig:fgremoval_result_Bmod2} displays the results for beam Model-II \rev{with a noise level of $\varepsilon=1.6\%$.  The higher noise amplitude compared to Model-I is due to the significantly reduced frequency dependence of the noise in Model-II, as evidenced in Figure \ref{fig: Beam_Fluctuation}.}
A notable distinction from Figure \ref{fig:fgremoval_result_Brand} is the increased over-subtraction at larger scales ($k\lesssim 0.15 \mathrm{h/Mpc}$), observed in both the traditional approach and the CBC method, regardless of the foreground model considered. This difference arises because, in this beam model, the two-dimensional noise is rescaled along the frequency direction with $\epsilon^d_{\mu}(\vu)=\epsilon^d(\mathbf{d} \nu/c)$, making the frequency dependence of the noise also scale-dependent. This behavior is apparent in Figure \ref{fig: Beam_Fluctuation}, where at lower $\vu$ values, corresponding to lower $k$ and shorter distance $\mathbf{d}$, the dependency on $\nu$ is less pronounced than at higher $k$ values. Such a condition leads to a unique interplay between the scales and the frequency of fluctuations. As a result, the SVD modes, derived from the frequency singular modes, do not perform uniformly across all scales. The maps produced are smoother in $\nu$ at larger angular scales, leading to over-removal in the SVD-cleaned maps. Despite this distinctive characteristic, our CBC method continues to outperform the traditional approach. As one can see, the level of improvement observed in Figure \ref{fig:fgremoval_result_Bmod2} is similar to  what is observed in Figure  \ref{fig:fgremoval_result_Brand}.

\rev{We also examine the performance of beam Model-II at various noise levels. The results are illustrated in Figures \ref{fig:fgremoval_result_Bmod2_eps08} and \ref{fig:fgremoval_result_Bmod2_eps24} for $\varepsilon = 0.8\%$ and $\varepsilon = 2.4\%$, respectively.  The overall feature of over-subtracted large scale modes persists in both situations. Similar to Model-I, the performance difference between CBC and the traditional method is much smaller for the low-noise beam model, with the improvement in the cross-correlation ratio ranging from ten percent to a few percent. Conversely, for a noise level of $\varepsilon = 2.4\%$,  the improvement reaches approximately one-third. 
}

\section{Discussion and Conclusions}

Foreground removal plays a pivotal role in the analysis of 21cm intensity mapping data. Most existing techniques require at least a basic understanding of the foreground, the instrument, or both. Even in blind methods, such as Principal Component Analysis (PCA), the specifics of the instrument significantly influence the efficiency of the results. Therefore, it is not surprising that these methods often show diminished performance in real-world applications, where actual instrumental beams are frequently noisy. Considerable efforts have been made towards accurate beam modeling and measurement. In this paper, we introduce a novel method that utilizing internal cross-correlation between different frequencies to calibrate the beam's variation with frequency.

To simulate the real-world beam fluctuations, we consider two specific models of beam noise. The first is a random noise in the three-dimensional $\vu\!-\!\nu$ space, while the second is a two-dimensional random noise in the instrumental space, which is then projected along the frequency axis with varying $\vu$ coordinates. Interestingly, the performances of these two models are somewhat distinct and complementary. Beam Model-II, in particular, experiences an over-subtraction at larger scales. However, when compared with traditional methods that overlook beam variation, our approach shows a considerable improvement. It achieves an approximate $50\%$ increase in the cross-correlation ratio and a significant reduction in the auto-power spectrum by several factors.

The filter constructed using our method incorporates two main factors: the relative change in the beam between two frequencies and the intrinsic temperature variation. To assess the impact of the later, we consider a simplified foreground model where the spatial pattern of the foreground remains constant. Interestingly, even substantial spatial fluctuations in the foreground temperature factor $F^T_{\nuref\nu}$ have a minimal effect on the performance. This resilience can be attributed to the smooth frequency variation of the foreground. Therefore, these findings further validate the effectiveness of our method.

\section*{Acknowledgments}
X.W. is supported by the National SKA Program of China (Grants Nos. 2022SKA0110200 and 2022SKA0110202). U-L.P. receives support from Ontario Research Fund—research Excellence Program (ORF-RE), Natural Sciences and Engineering Research Council of Canada (NSERC) [funding reference number RGPIN-2019-067, CRD 523638-201, 555585-20], Canadian Institute for Advanced Research (CIFAR), Canadian Foundation for Innovation (CFI), the National Science Foundation of China (Grants No. 11929301), Simons Foundation, Thoth Technology Inc, and Alexander von Humboldt Foundation.

\vspace{5mm}
%\facilities{HST(STIS), Swift(XRT and UVOT), AAVSO, CTIO:1.3m, CTIO:1.5m,CXO}

\software{CAMB~\citep{lewis2000efficient},
          NumPy~\citep{van2011numpy,harris2020array},
          SciPy~\citep{virtanen2020scipy},
          Astropy~\citep{robitaille2013astropy,price2018astropy,price2022astropy},
          HEALPix/healpy~\citep{2005ApJ...622..759G, Zonca2019}
          mpi4py~\citep{dalcin2021mpi4py},
          Matplotlib~\citep{hunter2007matplotlib}, 
          OpenMPI~\citep{gabriel2004open}
          }
          
\bibliography{ms_CBC}{}

\begin{thebibliography}{}
\expandafter\ifx\csname natexlab\endcsname\relax\def\natexlab#1{#1}\fi
\providecommand{\url}[1]{\href{#1}{#1}}
\providecommand{\dodoi}[1]{doi:~\href{http://doi.org/#1}{\nolinkurl{#1}}}
\providecommand{\doeprint}[1]{\href{http://ascl.net/#1}{\nolinkurl{http://ascl.net/#1}}}
\providecommand{\doarXiv}[1]{\href{https://arxiv.org/abs/#1}{\nolinkurl{https://arxiv.org/abs/#1}}}

\bibitem[{Abdurashidova {et~al.}(2022)Abdurashidova, Aguirre, Alexander, Ali,
  Balfour, Beardsley, Bernardi, Billings, Bowman, Bradley,
  {et~al.}}]{abdurashidova2022first}
Abdurashidova, Z., Aguirre, J.~E., Alexander, P., {et~al.} 2022, The
  Astrophysical Journal, 925, 221

\bibitem[{Alonso {et~al.}(2015)Alonso, Bull, Ferreira, \&
  Santos}]{alonso2015blind}
Alonso, D., Bull, P., Ferreira, P.~G., \& Santos, M.~G. 2015, Monthly Notices
  of the Royal Astronomical Society, 447, 400

\bibitem[{Alonso {et~al.}(2014)Alonso, Ferreira, \& Santos}]{alonso2014fast}
Alonso, D., Ferreira, P.~G., \& Santos, M.~G. 2014, Monthly Notices of the
  Royal Astronomical Society, 444, 3183

\bibitem[{Anderson {et~al.}(2014)Anderson, Aubourg, Bailey, Beutler, Bhardwaj,
  Blanton, Bolton, Brinkmann, Brownstein, Burden,
  {et~al.}}]{anderson2014clustering}
Anderson, L., Aubourg, E., Bailey, S., {et~al.} 2014, Monthly Notices of the
  Royal Astronomical Society, 441, 24

\bibitem[{Astropy~Collaboration {et~al.}(2018)Astropy~Collaboration,
  Sip{\H{o}}cz, G{\"u}nther, Lim, Crawford, Conseil, Shupe, Craig, Dencheva,
  Ginsburg, {et~al.}}]{price2018astropy}
Astropy~Collaboration, Price-Whelan, A.~M., Sip{\H{o}}cz, B., G{\"u}nther, H.,
  {et~al.} 2018, The Astronomical Journal, 156, 123

\bibitem[{Astropy~Collaboration {et~al.}(2022)Astropy~Collaboration, Lim, Earl,
  Starkman, Bradley, Shupe, Patil, Corrales, Brasseur, N{\"o}the,
  {et~al.}}]{price2022astropy}
Astropy~Collaboration, Price-Whelan, A.~M., Lim, P.~L., Earl, N., {et~al.}
  2022, The Astrophysical Journal, 935, 167

\bibitem[{Astropy~Collaboration {et~al.}(2013)Astropy~Collaboration, Tollerud,
  Greenfield, Droettboom, Bray, Aldcroft, Davis, Ginsburg, Price-Whelan,
  Kerzendorf, {et~al.}}]{robitaille2013astropy}
Astropy~Collaboration, Robitaille, T.~P., Tollerud, E.~J., Greenfield, P.,
  {et~al.} 2013, Astronomy \& Astrophysics, 558, A33

\bibitem[{B\'egin {et~al.}(2022)B\'egin, Liu, \& Gorce}]{PhysRevD.105.083503}
B\'egin, J.-M., Liu, A., \& Gorce, A. 2022, Phys. Rev. D, 105, 083503,
  \dodoi{10.1103/PhysRevD.105.083503}

\bibitem[{{Bernardi} {et~al.}(2011){Bernardi}, {Mitchell}, {Ord}, {Greenhill},
  {Pindor}, {Wayth}, \& {Wyithe}}]{Bernardi_2011MNRAS}
{Bernardi}, G., {Mitchell}, D.~A., {Ord}, S.~M., {et~al.} 2011, Monthly Notices
  of the Royal Astronomical Society, 413, 411,
  \dodoi{10.1111/j.1365-2966.2010.18145.x}

\bibitem[{Bigot-Sazy {et~al.}(2015)Bigot-Sazy, Dickinson, Battye, Browne, Ma,
  Maffei, Noviello, Remazeilles, \& Wilkinson}]{bigot2015simulations}
Bigot-Sazy, M.-A., Dickinson, C., Battye, R.~A., {et~al.} 2015, Monthly Notices
  of the Royal Astronomical Society, 454, 3240

\bibitem[{Chang {et~al.}(2015)Chang, Monstein, Refregier, Amara, Glauser, \&
  Casura}]{chang2015beam}
Chang, C., Monstein, C., Refregier, A., {et~al.} 2015, Publications of the
  Astronomical Society of the Pacific, 127, 1131

\bibitem[{Chapman {et~al.}(2012)Chapman, Abdalla, Harker, Jeli{\'c},
  Labropoulos, Zaroubi, Brentjens, de~Bruyn, \&
  Koopmans}]{chapman2012foreground}
Chapman, E., Abdalla, F.~B., Harker, G., {et~al.} 2012, Monthly Notices of the
  Royal Astronomical Society, 423, 2518

\bibitem[{{Coles} \& {Jones}(1991)}]{Coles1991MNRAS}
{Coles}, P., \& {Jones}, B. 1991, Monthly Notices of the Royal Astronomical
  Society, 248, 1, \dodoi{10.1093/mnras/248.1.1}

\bibitem[{Collaboration {et~al.}(2023)Collaboration, Adame, Aguilar, Ahlen,
  Alam, Aldering, Alexander, Alfarsy, Prieto, Alvarez,
  {et~al.}}]{collaboration2023early}
Collaboration, D., Adame, A., Aguilar, J., {et~al.} 2023, arXiv preprint
  arXiv:2306.06308

\bibitem[{Collaboration: {et~al.}(2016)Collaboration:, Abbott, Abdalla,
  Aleksi{\'c}, Allam, Amara, Bacon, Balbinot, Banerji, Bechtol,
  {et~al.}}]{dark2016dark}
Collaboration:, D. E.~S., Abbott, T., Abdalla, F., {et~al.} 2016, Monthly
  Notices of the Royal Astronomical Society, 460, 1270

\bibitem[{Colless {et~al.}(2001)Colless, Dalton, Maddox, Sutherland, Norberg,
  Cole, Bland-Hawthorn, Bridges, Cannon, Collins, {et~al.}}]{colless20012df}
Colless, M., Dalton, G., Maddox, S., {et~al.} 2001, Monthly Notices of the
  Royal Astronomical Society, 328, 1039

\bibitem[{Condon \& Ransom(2016)}]{condon2016essential}
Condon, J.~J., \& Ransom, S.~M. 2016, Essential radio astronomy, Vol.~2
  (Princeton University Press)

\bibitem[{Cunnington {et~al.}(2023)Cunnington, Li, Santos, Wang, Carucci,
  Irfan, Pourtsidou, Spinelli, Wolz, Soares, {et~al.}}]{cunnington2023h}
Cunnington, S., Li, Y., Santos, M.~G., {et~al.} 2023, Monthly Notices of the
  Royal Astronomical Society, 518, 6262

\bibitem[{Dalcin \& Fang(2021)}]{dalcin2021mpi4py}
Dalcin, L., \& Fang, Y.-L.~L. 2021, Computing in Science \& Engineering, 23, 47

\bibitem[{DeBoer {et~al.}(2017)DeBoer, Parsons, Aguirre, Alexander, Ali,
  Beardsley, Bernardi, Bowman, Bradley, Carilli, {et~al.}}]{deboer2017hydrogen}
DeBoer, D.~R., Parsons, A.~R., Aguirre, J.~E., {et~al.} 2017, Publications of
  the Astronomical Society of the Pacific, 129, 045001

\bibitem[{Delabrouille {et~al.}(2013)Delabrouille, Betoule, Melin,
  Miville-Desch{\^e}nes, Gonzalez-Nuevo, Le~Jeune, Castex, De~Zotti, Basak,
  Ashdown, {et~al.}}]{delabrouille2013pre}
Delabrouille, J., Betoule, M., Melin, J.-B., {et~al.} 2013, Astronomy \&
  Astrophysics, 553, A96

\bibitem[{Drinkwater {et~al.}(2010)Drinkwater, Jurek, Blake, Woods, Pimbblet,
  Glazebrook, Sharp, Pracy, Brough, Colless, {et~al.}}]{drinkwater2010wigglez}
Drinkwater, M.~J., Jurek, R.~J., Blake, C., {et~al.} 2010, Monthly Notices of
  the Royal Astronomical Society, 401, 1429

\bibitem[{Gabriel {et~al.}(2004)Gabriel, Fagg, Bosilca, Angskun, Dongarra,
  Squyres, Sahay, Kambadur, Barrett, Lumsdaine, {et~al.}}]{gabriel2004open}
Gabriel, E., Fagg, G.~E., Bosilca, G., {et~al.} 2004, in Recent Advances in
  Parallel Virtual Machine and Message Passing Interface: 11th European PVM/MPI
  Users’ Group Meeting Budapest, Hungary, September 19-22, 2004. Proceedings
  11, Springer, 97--104

\bibitem[{Gheller \& Vazza(2022)}]{gheller2022convolutional}
Gheller, C., \& Vazza, F. 2022, Monthly Notices of the Royal Astronomical
  Society, 509, 990

\bibitem[{Gillet {et~al.}(2019)Gillet, Mesinger, Greig, Liu, \&
  Ucci}]{gillet2019deep}
Gillet, N., Mesinger, A., Greig, B., Liu, A., \& Ucci, G. 2019, Monthly Notices
  of the Royal Astronomical Society, 484, 282

\bibitem[{{G{\'o}rski} {et~al.}(2005){G{\'o}rski}, {Hivon}, {Banday},
  {Wandelt}, {Hansen}, {Reinecke}, \& {Bartelmann}}]{2005ApJ...622..759G}
{G{\'o}rski}, K.~M., {Hivon}, E., {Banday}, A.~J., {et~al.} 2005, \apj, 622,
  759, \dodoi{10.1086/427976}

\bibitem[{Harris {et~al.}(2020)Harris, Millman, Van Der~Walt, Gommers,
  Virtanen, Cournapeau, Wieser, Taylor, Berg, Smith,
  {et~al.}}]{harris2020array}
Harris, C.~R., Millman, K.~J., Van Der~Walt, S.~J., {et~al.} 2020, Nature, 585,
  357

\bibitem[{Haslam {et~al.}(1982)Haslam, Salter, Stoffel, \&
  Wilson}]{haslam1982408}
Haslam, C., Salter, C., Stoffel, H., \& Wilson, W. 1982, Astronomy and
  Astrophysics Supplement Series, 47, 1

\bibitem[{{Hu} {et~al.}(2021){Hu}, {Li}, {Wang}, {Wu}, {Zhang}, {Zhu}, {Zuo},
  {Lagache}, {Ma}, {Santos}, \& {Chen}}]{Hu2021MNRAS}
{Hu}, W., {Li}, Y., {Wang}, Y., {et~al.} 2021, MNRAS, 508, 2897,
  \dodoi{10.1093/mnras/stab2728}

\bibitem[{Hunter(2007)}]{hunter2007matplotlib}
Hunter, J.~D. 2007, Computing in science \& engineering, 9, 90

\bibitem[{Hunter {et~al.}(2011)Hunter, Schwab, White, Ford, Ghigo, Maddalena,
  Mason, Nelson, Prestage, Ray, {et~al.}}]{hunter2011holographic}
Hunter, T.~R., Schwab, F.~R., White, S.~D., {et~al.} 2011, Publications of the
  Astronomical Society of the Pacific, 123, 1087

\bibitem[{Iheanetu {et~al.}(2019)Iheanetu, Girard, Smirnov, Asad, de~Villiers,
  Thorat, Makhathini, \& Perley}]{iheanetu2019primary}
Iheanetu, K., Girard, J., Smirnov, O., {et~al.} 2019, Monthly Notices of the
  Royal Astronomical Society, 485, 4107

\bibitem[{Ivezi{\'c} {et~al.}(2019)Ivezi{\'c}, Kahn, Tyson, Abel, Acosta,
  Allsman, Alonso, AlSayyad, Anderson, Andrew, {et~al.}}]{ivezic2019lsst}
Ivezi{\'c}, {\v{Z}}., Kahn, S.~M., Tyson, J.~A., {et~al.} 2019, The
  Astrophysical Journal, 873, 111

\bibitem[{Jones {et~al.}(2004)Jones, Saunders, Colless, Read, Parker, Watson,
  Campbell, Burkey, Mauch, Moore, {et~al.}}]{jones20046df}
Jones, D.~H., Saunders, W., Colless, M., {et~al.} 2004, Monthly Notices of the
  Royal Astronomical Society, 355, 747

\bibitem[{La~Plante \& Ntampaka(2019)}]{la2019machine}
La~Plante, P., \& Ntampaka, M. 2019, The Astrophysical Journal, 880, 110

\bibitem[{Lewis {et~al.}(2000)Lewis, Challinor, \&
  Lasenby}]{lewis2000efficient}
Lewis, A., Challinor, A., \& Lasenby, A. 2000, The Astrophysical Journal, 538,
  473

\bibitem[{Li {et~al.}(2023)Li, Wang, Deng, Yang, Hu, Liu, Zhao, Zuo, Shu, Li,
  {et~al.}}]{li2023fast}
Li, Y., Wang, Y., Deng, F., {et~al.} 2023, arXiv preprint arXiv:2305.06405

\bibitem[{{Liao} {et~al.}(2016){Liao}, {Chang}, {Kuo}, {Masui}, {Oppermann},
  {Pen}, \& {Peterson}}]{Liao2016ApJ}
{Liao}, Y.-W., {Chang}, T.-C., {Kuo}, C.-Y., {et~al.} 2016, \apj, 833, 289,
  \dodoi{10.3847/1538-4357/833/2/289}

\bibitem[{Liao {et~al.}(2016)Liao, Chang, Kuo, Masui, Oppermann, Pen, \&
  Peterson}]{Liao_2016}
Liao, Y.-W., Chang, T.-C., Kuo, C.-Y., {et~al.} 2016, The Astrophysical
  Journal, 833, 289, \dodoi{10.3847/1538-4357/833/2/289}

\bibitem[{Makinen {et~al.}(2021)Makinen, Lancaster, Villaescusa-Navarro,
  Melchior, Ho, Perreault-Levasseur, \& Spergel}]{makinen2021deep21}
Makinen, T.~L., Lancaster, L., Villaescusa-Navarro, F., {et~al.} 2021, Journal
  of Cosmology and Astroparticle Physics, 2021, 081

\bibitem[{Mangena {et~al.}(2020)Mangena, Hassan, \&
  Santos}]{mangena2020constraining}
Mangena, T., Hassan, S., \& Santos, M.~G. 2020, Monthly Notices of the Royal
  Astronomical Society, 494, 600

\bibitem[{Masui {et~al.}(2013)Masui, Switzer, Banavar, Bandura, Blake, Calin,
  Chang, Chen, Li, Liao, {et~al.}}]{masui2013measurement}
Masui, K., Switzer, E., Banavar, N., {et~al.} 2013, The Astrophysical Journal
  Letters, 763, L20

\bibitem[{Ni {et~al.}(2022)Ni, Li, Gao, \& Zhang}]{ni2022eliminating}
Ni, S., Li, Y., Gao, L.-Y., \& Zhang, X. 2022, The Astrophysical Journal, 934,
  83

\bibitem[{Paul {et~al.}(2023)Paul, Santos, Chen, \& Wolz}]{paul2023first}
Paul, S., Santos, M.~G., Chen, Z., \& Wolz, L. 2023, arXiv preprint
  arXiv:2301.11943

\bibitem[{Santos {et~al.}(2005)Santos, Cooray, \&
  Knox}]{santos2005multifrequency}
Santos, M.~G., Cooray, A., \& Knox, L. 2005, The Astrophysical Journal, 625,
  575

\bibitem[{Scott \& Rees(1990)}]{scott199021}
Scott, D., \& Rees, M.~J. 1990, Monthly Notices of the Royal Astronomical
  Society, vol. 247, p. 510, 247, 510

\bibitem[{{Shaw} {et~al.}(2014){Shaw}, {Sigurdson}, {Pen}, {Stebbins}, \&
  {Sitwell}}]{Shaw_2014ApJ}
{Shaw}, J.~R., {Sigurdson}, K., {Pen}, U.-L., {Stebbins}, A., \& {Sitwell}, M.
  2014, \apj, 781, 57, \dodoi{10.1088/0004-637X/781/2/57}

\bibitem[{Shi {et~al.}(2023)Shi, Chang, Zhang, Shan, Zhang, Zhou, Jiang, \&
  Wang}]{shi202321}
Shi, F., Chang, H., Zhang, L., {et~al.} 2023, arXiv preprint arXiv:2310.06518

\bibitem[{Stone(2002)}]{stone2002independent}
Stone, J.~V. 2002, Trends in cognitive sciences, 6, 59

\bibitem[{{Sullivan} {et~al.}(2012){Sullivan}, {Morales}, {Hazelton}, {Arcus},
  {Barnes}, {Bernardi}, {Briggs}, {Bowman}, {Bunton}, {Cappallo}, {Corey},
  {Deshpande}, {deSouza}, {Emrich}, {Gaensler}, {Goeke}, {Greenhill}, {Herne},
  {Hewitt}, {Johnston-Hollitt}, {Kaplan}, {Kasper}, {Kincaid}, {Koenig},
  {Kratzenberg}, {Lonsdale}, {Lynch}, {McWhirter}, {Mitchell}, {Morgan},
  {Oberoi}, {Ord}, {Pathikulangara}, {Prabu}, {Remillard}, {Rogers}, {Roshi},
  {Salah}, {Sault}, {Udaya Shankar}, {Srivani}, {Stevens}, {Subrahmanyan},
  {Tingay}, {Wayth}, {Waterson}, {Webster}, {Whitney}, {Williams}, {Williams},
  \& {Wyithe}}]{Sullivan2012ApJ}
{Sullivan}, I.~S., {Morales}, M.~F., {Hazelton}, B.~J., {et~al.} 2012, \apj,
  759, 17, \dodoi{10.1088/0004-637X/759/1/17}

\bibitem[{Switzer {et~al.}(2013)Switzer, Masui, Bandura, Calin, Chang, Chen,
  Li, Liao, Natarajan, Pen, {et~al.}}]{switzer2013determination}
Switzer, E., Masui, K., Bandura, K., {et~al.} 2013, Monthly Notices of the
  Royal Astronomical Society: Letters, 434, L46

\bibitem[{Switzer {et~al.}(2015)Switzer, Chang, Masui, Pen, \&
  Voytek}]{switzer2015interpreting}
Switzer, E.~R., Chang, T.-C., Masui, K.~W., Pen, U.-L., \& Voytek, T.~C. 2015,
  The Astrophysical Journal, 815, 51

\bibitem[{Tharwat(2021)}]{tharwat2021independent}
Tharwat, A. 2021, Applied Computing and Informatics, 17, 222

\bibitem[{Van Der~Walt {et~al.}(2011)Van Der~Walt, Colbert, \&
  Varoquaux}]{van2011numpy}
Van Der~Walt, S., Colbert, S.~C., \& Varoquaux, G. 2011, Computing in science
  \& engineering, 13, 22

\bibitem[{Villanueva-Domingo \&
  Villaescusa-Navarro(2021)}]{villanueva2021removing}
Villanueva-Domingo, P., \& Villaescusa-Navarro, F. 2021, The Astrophysical
  Journal, 907, 44

\bibitem[{Virtanen {et~al.}(2020)Virtanen, Gommers, Oliphant, Haberland, Reddy,
  Cournapeau, Burovski, Peterson, Weckesser, Bright,
  {et~al.}}]{virtanen2020scipy}
Virtanen, P., Gommers, R., Oliphant, T.~E., {et~al.} 2020, Nature methods, 17,
  261

\bibitem[{Wadekar {et~al.}(2021)Wadekar, Villaescusa-Navarro, Ho, \&
  Perreault-Levasseur}]{wadekar2021hinet}
Wadekar, D., Villaescusa-Navarro, F., Ho, S., \& Perreault-Levasseur, L. 2021,
  The Astrophysical Journal, 916, 42

\bibitem[{Wolz {et~al.}(2014)Wolz, Abdalla, Blake, Shaw, Chapman, \&
  Rawlings}]{wolz2014effect}
Wolz, L., Abdalla, F., Blake, C., {et~al.} 2014, Monthly Notices of the Royal
  Astronomical Society, 441, 3271

\bibitem[{Wolz {et~al.}(2022)Wolz, Pourtsidou, Masui, Chang, Bautista,
  M{\"u}ller, Avila, Bacon, Percival, Cunnington, {et~al.}}]{wolz2022h}
Wolz, L., Pourtsidou, A., Masui, K.~W., {et~al.} 2022, Monthly Notices of the
  Royal Astronomical Society, 510, 3495

\bibitem[{Zhang {et~al.}(2019)Zhang, Wu, Li, Kr{\v{c}}o, Staveley-Smith, Tang,
  Qian, Liu, Jin, Yue, {et~al.}}]{zhang2019status}
Zhang, K., Wu, J., Li, D., {et~al.} 2019, Science China Physics, Mechanics \&
  Astronomy, 62, 1

\bibitem[{Zonca {et~al.}(2019)Zonca, Singer, Lenz, Reinecke, Rosset, Hivon, \&
  Gorski}]{Zonca2019}
Zonca, A., Singer, L., Lenz, D., {et~al.} 2019, Journal of Open Source
  Software, 4, 1298, \dodoi{10.21105/joss.01298}

\bibitem[{{Zuo} {et~al.}(2023){Zuo}, {Chen}, \& {Mao}}]{Zuo2023ApJ}
{Zuo}, S., {Chen}, X., \& {Mao}, Y. 2023, \apj, 945, 38,
  \dodoi{10.3847/1538-4357/acb822}

\end{thebibliography}
\bibliographystyle{aasjournal}

\end{document}